\input epsf
\documentclass[preprint,showpacs,preprintnumbers,amsmath,amssymb]{revtex4}

\usepackage{epsfig}
\usepackage{graphicx}
\usepackage{dcolumn}
\usepackage{bm}

\begin{document}
\preprint{FU-Berlin/2004}
\title{
Canonical local algorithms for spin systems: Heat Bath and Hasting's
methods
       }
\author{
D. Loison$^1$\footnote{corresponding author: Damien.Loison@physik.fu-berlin.de}%
\footnote{Work supported by Deutsche Forschungsgemeinschaft under Grant No.\ Scho158/10},
C. Qin$^2$, 
K.D. Schotte$^1$ and 
X.F. Jin$^3$ }
\affiliation{
$^1$Institut f\"ur Theoretische Physik, Freie Universit\"at Berlin,
Arnimallee 14, 14195 Berlin, Germany\\
$^2$Institute for Materials Research, Tohoku University, 
Sendai 980-8577, Japan\\
$^3$Surface Physics Laboratory, Fudan University, Shanghai 200433, China
}

\begin{abstract}
We introduce new fast canonical local 
algorithms for discrete and continuous spin systems.
We show that for a broad selection of spin systems
they compare favorably to the known ones except for 
the Ising $\pm1$ spins. 
The new procedures use discretization scheme and the 
necessary information have to be stored 
in computer memory before the simulation.
The models for testing discrete spins are the
Ising $\pm1$, the general Ising $S$ or Blume-Capel model,
the Potts and the clock models.
The continuous spins we examine are the $O(N)$ models, including 
the continuous Ising model ($N=1$), the $\phi^4$ Ising model ($N=1$), 
the $XY$ model ($N=2$), the Heisenberg model ($N=3$), the
$\phi^4$ Heisenberg model ($N=3$), the $O(4)$ model with applications
to the $SU(2)$ lattice gauge theory, and the general $O(N)$ vector spins 
with $N\ge5$.
\end{abstract}

\vspace{1.cm}
P.A.C.S. numbers:05.70.Fh, 64.60.Cn, 75.10.Hk, 75.10.Nr
\vspace{1.cm}

\maketitle


\section{INTRODUCTION}

Spin systems are one of the most studied subjects in physics 
for their own interest but also because
many other problems can be mapped on them. 
Since few exact calculations are available, numerical Monte 
Carlo simulations are extensively used to study them. 
One of the most popular method is the Metropolis 
algorithm\cite{Metropolis}, rightly because of its simplicity and
its general applicability. 
The Metropolis algorithm is the easiest one to implement since it 
uses no prior informations and
automatically matches the properties of the systems.
However, using more of the available informations 
one should obtain algorithms better suited to spin systems.
The algorithms proposed in this article follow this strategy at 
the price that they are slightly more complicated.
To help implementations all programs discussed in this
article are accessible at our homepage \cite{homepage}.

The algorithms studied in this article use only local updates.
The non-local cluster algorithms are certainly better suited to study
ferromagnetic second order phase transitions but  cannot
be used for frustrated spin systems and for first order transitions.
Moreover, even when available, cluster algorithms
should  be used in combination with local algorithm \cite{Plascak}. 
The gain in efficiency quoted in the present article must
be understood as the gain for the part corresponding to the local
algorithm alone.
Further we have checked the influence on the gain of other algorithms 
like the exchange algorithm\cite{exchange}, the over-relaxation algorithm
\cite{Overrelaxation} or the multicanonical algorithm \cite{Berg}.
We think that, if the multi-spins coding cannot be implemented,
the methods introduced in this article are the fastest 
possible local updates and no real improvements could be made.
If the multi-spins coding can be implemented
it should be used because the gain will be better.
However, this method is less general than the methods
introduced in this article and moreover 
exist only for discrete spins.

Simulations to test the algorithms have been done for several lattices
with frustrated or non frustrated spin systems, using small lattice sizes $L$. 
The examples studied are the
two-dimensional ferromagnetic square lattice (2s),
the two-dimensional antiferromagnetic triangular lattice (2t),
the three-dimensional antiferromagnetic stacked triangular lattice (3t),
the three-dimensional ferromagnetic and anti--ferromagnetic
cubic lattice (3c \& 3ca), and the four-dimensional spin glasses
on a cubic lattice with random $\pm J$ interaction (4sg).

The comparative efficiency of the algorithm
studied does not depend on the size,
due to the local nature of the updates.
For a second order transition the integrated autocorrelation time
(see for example the appendix of \cite{Loison_Simon}) follows
the same law $\tau = \tau_0 \cdot L^z$ with $z\approx2$ and only the value of
$\tau_0$  depends on the algorithm used. 
In a strong first order transition $\tau$ has an exponential form
for a local algorithm but 
the multicanonical algorithm in combination gives  
a power law behavior with a different $z$ \cite{Berg}.
Also, because algorithms are local, the type of 
lattices and interactions do not have a strong influence on the 
comparative efficiency of the algorithms.
The important parameter is the temperature or more precisely 
the local field, that is the sum of the
neighboring spins associated with interactions divided by the temperature.
Therefore the methods introduced
are not restricted to the lattices studied here
but will be applicable to other lattices as well.

Most of the  algorithms proposed use
arrays to store values calculated prior to the simulation. 
The performances of the programs depend therefore crucially
on the access
time to the memory and in particular on the size of cache memory.
For the simulations an Athlon 1800 processor
with $250\,$Kb cache
memory has been used. This amount has quadrupled to $1\,$Mb with new 
processors now available (in 2004)
and therefore the performances of the proposed algorithms  
should also increase.
Moreover the performance also strongly depends on the compiler
used. We used the Intel compiler ``icc'' 
with the optimization ``-O''
and obtained a code two times faster than the one with the 
``gcc'' compiler.

In the following we start with a reexamination
of the detailed balance condition.
Then we introduce different algorithms trying them out on
several types of 
discrete and continuous spins. 
The discrete spin systems we consider are
the $\pm1$ Ising spin, 
the general Ising spin with S $> 2$ states, the Potts and the clock models.
For the continuous spins we investigate
continuous Ising spins $-1 \geq S \geq +1$, the Ising $\phi^4$--model,
$XY$ spins, Heisenberg spins, the Heisenberg $\phi^4$--model,
and also the $O(N)$ spin models for $N\geq4$.


\section{Detailed Balance condition}

The basis in the simulation techniques we are going to discuss is
the detailed balance in the updating process\cite{Gentle}.
There the essential point is symmetry in time: 
the probability $P(A)$ of an initial configuration $A$ multiplied
by the transition
probability $T_{A \rightarrow B}$ to a new or final configuration
$B$ should be equal to the
probability for the inverse process  starting from the configuration
$B$ with $P(B)$ and
transition probability $T_{B \rightarrow A}$, so that
\begin{eqnarray}
\label{detailed_balance_ini}
P(A) \cdot T_{A\rightarrow B} =
P(B) \cdot T_{B\rightarrow A} \ .
\end{eqnarray}

In choosing the transition probabilities $T$ properly the spin
configurations will
be visited with the weights $P$ and these are for a canonical system 
the Boltzmann weights.
To be more flexible the detailed balance condition (1) should be
rewritten in the form
\begin{eqnarray}
\label{detailed_balance}
\frac{P(A)}{f(B,A)}\, f(B,A) \cdot T_{A\rightarrow B}\, =\,
\frac{P(B)}{f(A,B)}\, f(A,B) \cdot T_{B\rightarrow A} 
\end{eqnarray}
with an auxiliary function $f$ depending on the two states $A$ and $B$.
There are essentially two ways to satisfy these equations,
the ``heat bath'' method and the ``Hasting'' method, this last one
includes the Metropolis method.

In the first case the transition probability $T_{A\rightarrow B} = T(B)$
and the function $f(A,\,B) = f(B)$ depend only on the final state $B$,
so that the detailed balance condition is reduced to
\begin{equation}
\label{heat-bath}
T_{A\rightarrow B} = P(B) = \,\frac{P(B)}{f(B)}\cdot f(B) \ .
\end{equation}
The simplified form $T(B)=P(B)$ should be used if the
probability can be integrated ``easily'' and inverted like for the 
discrete spins or Heisenberg spins.
In principle any probability function $P(B)$ depending on one variable
could be integrated and inverted numerically and the result stored in tables.
However, this method is not recommended for spin systems like $XY$ spins
for example, since the tables become too large and the algorithm will be 
costly in computing time to get sufficiently small errors.
For such a case it is better to use the second form of eq.~(\ref{heat-bath})
and proceed as follows.

\def\ran{\hbox{\rm ran\thinspace}}
The first procedure is the ``heat bath'' method.
One finds a new configuration $B=F^{-1}(\ran)$ using a random number
``\ran'' and the inverse of the function $F$ which is the sum or the integral of $f$. 
Having found the new state $B$, it
is accepted with the probability $\frac{P(B)}{f(B)}$ with a new random
number $0\leq \ran_2 < 1$ by the usual rejection method if
$\frac{P(B)}{f(B)}\geq \ran_2$. The function $f(B)$ must be always larger
than $P(B)$ but the difference should be small since otherwise too many
choices for the new states will be rejected. 
The simplified case corresponds to $f(B)=P(B)$ and no second random number
is needed.

The second procedure is Hasting's method \cite{Hasting} with
the transition probability equal to
\begin{eqnarray}
\label{Hasting}
T_{A\rightarrow B}=f(A,B) \cdot 
                  \min \left( \frac{P(B)\cdot f(B,A)}{P(A)\cdot
				  f(A,B)}\,,\,1\right) \ .
\end{eqnarray}
Putting this into
(\ref{detailed_balance}) it is not difficult
to see that the detailed balance condition is indeed fulfilled.
The function $f$ can here be smaller or larger than $P$ which is
of advantage compared to the stricter condition for the 
heat bath method.
If one chooses $f(A,B)=1$ for any $A$ and $B$ one gets the transition
probability of the Metropolis 
method \cite{Metropolis}.
The gain in efficiency due to Hasting's generalization can be quite
large in comparison to the Metropolis update
depending on the spin model, the choice of the function $f$ and
the temperature range of the simulation.

Whatever the method used will be, it should be ergodic, i.e.\ all
configurations should be accessible. Since our special choices for the function $f$ 
will be of a kind for which the local updates do not change in an essential
way from the ones in use, we will not come back to this point.
In the following we have to treat discrete and continuous spin systems
separately.


\section{Discrete model}

\subsection{Simulation methods}

We present in this section six algorithms we have implemented
for discrete spins.
We assume that the spins have $q$ possible states (1, 2, $\cdots$ $q$).
The actual or old state is $A$ and the new state $B$ as before.

\subsubsection{Metropolis}
\noindent
The first method one tries usually is the Metropolis
algorithm \cite{Metropolis} abbreviated as  {\sl Me} in the figures. 
As discussed in the preceding section in eq.~(\ref{Hasting}) the choice
is $f(A,\,B)=1$ whatever the states $A,\,B$ are. The 
implementation is as follows:
\begin{itemize}
\item[1.] Choose randomly one of $q$ states $B$ using
$\,{\rm int}(q\cdot\ran_1)$ with
$\ran_1$ a random number in the range $0\leq \ran_1 < 1$ or in the
interval $[0,1[\,$. The function ``int''
gives the largest integer smaller than the floating point number.
\item[2.] Accept this new state with another random number,
if $P(B)/P(A)> \ran_2$.
\end{itemize}
There are two problems to be noted: 
\begin{itemize}
\item[a.] The choice of the new state can be identical to the old one.
\item[b.] The choice of a new state is equiprobable, but in particular at low
temperatures the Boltzmann probabilities $P(B)$ will be quite different.
This fact will lead to a large and unproductive rejection rate.
\end{itemize}

\subsubsection{Restricted Metropolis}
\noindent
A way to solve the first problem is to 
choose the new state randomly omitting 
the actual state. We denote this  as the restricted Metropolis 
algorithm or {\sl Me}$_\delta$.
With the choice $f(A,B\! \neq\! A)=1$ and $f(A,A)=0\,$ 
eq.~(\ref{Hasting}) is practically the same as for the Metropolis
algorithm {\sl Me}. The implementation is therefore:
\begin{itemize}
\item[1.] Choose randomly one state $B$ using 
$\,{\rm int}((q\! - \! 1)\cdot \ran_1)\,$ and use a list to find
randomly one of the $q \! - \!1$ states different from $A$.
\item[2.] Accept this new state if $P(B)/P(A)> \ran_2\,$.
\end{itemize}
Since the actual state is no longer tested 
using {\sl Me}$_\delta$ one gains compared to {\sl Me} an amount $1/q$
in the updating process. If $q$ is large, this difference becomes negligible.
For the Ising two-states case only the opposite state is tested
anyway. But as a standard for $q > 2$ it is not used, as far as we know,
where also the performance is improved in a measurable way  as we will show.

As to the second problem b described at the end of the previous section
it will be addressed by the heat bath method discussed next.
In addition a restricted heat bath 
will solve both problems a and b simultaneously.

\subsubsection{Direct Heat Bath \label{DHB_discrete}}
\noindent
In discrete system the number of states is finite and one can simply 
choose the function $f=P$ in eq.~(\ref{heat-bath}). We call
it here Direct Heat Bath or abbreviated {\sl  DHB}. The implementation is simple:
\begin{itemize}
\item[1.] Calculate the  $\hbox{\sl Norm} = P(0)+P(1) + \cdots$.
\item[2.] Take a random number \ran\/ in the range $[0,1[\,$.
\item[3.] If $\ran <P(1)/\hbox{\sl Norm}$, the new configuration is chosen as $1$.
\item[4.] Else if $\ran <(P(1)+P(2))/\hbox{\sl Norm}$, the new configuration
          is chosen as $2$.
\item[5.] and so forth \dots 
\end{itemize}
This algorithm has two flaws:
\begin{itemize}
\item[a.] The new state can be identical to the old one.
\item[b.] If there are too many possible states,
          the algorithm becomes very slow.
\end{itemize}
The first problem is similar to the Metropolis algorithm
but at least the new state is chosen among really accessible states
according to  their probability $P(B)$. Therefore the acceptance
rate is 1.

This algorithm is always more efficient than 
the Metropolis algorithm {\sl Me} and also  
the restricted Metropolis method {\sl Me}$_{\delta}\,$.
Only for Ising spins $S = \pm1$ the $Me_\delta$ method is
more efficient than this heat bath method. 

\subsubsection{Direct Heat Bath with Walker's alias \label{section_Walker}}
\noindent
To solve the problem of the inefficiency of the Heat Bath method for a larger
number of states $q$ we propose to use Walker's alias method \cite{Walker}. 
Since this procedure was, to our knowledge, never used
in spin simulations, it needs some explanations.

Walker's alias method handles in an economic way which new state to choose
among the $q$ possibilities. The probabilities $P(q)$ for a new state $i$ are not piled
up on top of each other to sum up to 1 as in the simple heat bath described
before, but stored in $q$ different boxes of equal height $1/q$.
Walker's construction has in each box only one or two different
probabilities. For an example with $q=3$ see fig.~\ref{Clock_a}(b).
Before the simulation starts one must have calculated and stored the
probabilities $P^{i}_{limit}$ which divides each box $i$. 
The upper states in each box must also be stored in an array. 
These states as ``subtenants''  have an ``alias'' whereas the lower ones have
the box number as correct address for the state $i$.

\noindent
The implementation has the following steps: 
\begin{itemize}
\item[1.] Choose the box $i$ using  $i = {\rm int}(q\cdot \ran) + \!1\,$
with a random number $\ran \in [0,1[ \,$.
\item[2.] If $\ran < P^{i}_{limit}$, choose the new state as $i$,
          otherwise take the new state as Alias$[i]\,$.
\end{itemize}
The consumption time is therefore independent on the number of
states. The only limitation is the memory needed to store the arrays. 
The method to generate the arrays can be found in the ref.~\cite{Peterson}
and the implementation at our homepage \cite{homepage}.
The gain, using this method in comparison to the standard
heat bath, is from one to ten depending on the number of states.
In the figures when using this implementation of the heat bath algorithm,
we denote it shortly as {\sl  AW}  for Alias Walker. 

\subsubsection{Restricted Direct Heat Bath}
\noindent
We want now to improve the heat bath algorithm by avoiding
to choose the new state identical to the old one.
The old state is $A$, the chosen new state $B$ and the other
accessible states $C$. 
To find the form of the function $f$ in eq.~(\ref{Hasting}) we start
with the heat bath method and get $f(A,B)=\frac{P(B)}{P(B)+P(C)} \>$.
The denominator would be 1 and $f(A,\,B) = P(B)$ if $P(A)$ were included
in the sum $P(B) \! + \!P(C)$. For the reverse process starting from $B$ and
choosing the state $A$ in the accessible states $A+C$, we have
$f(B,A)=\frac{P(A)}{P(A)+P(C)} \>$.

\noindent The implementation of the algorithm is therefore:
\begin{itemize}
\item[1.] Choose $B$ in $B+C$ using heat bath method, see above.
\item[2.] Accept $B$ if $\, \frac{P(B)+P(C)}{P(A)+P(C)}\,\geq \ran$ 
according to eq.~(\ref{Hasting}).
\end{itemize}
We call this algorithm the restricted heat bath or {\sl  DHB}$_{\delta}\,$.
For two states as the Ising $\pm1$ spins,
this algorithm is equivalent to the restricted Metropolis one {\sl Me}$_{\delta}\,$.

\subsubsection{Restricted Direct Heat Bath with Walker's Alias}
\noindent
Finally, as above we make 
use of the Walker algorithm to accelerate the simulation.
We call this algorithm
the restricted Alias Walker Hasting Algorithm or abbreviated 
{\sl  AWH}$_{\delta}\,$. The arrays to store the values calculated before
the simulation are $q$ times 
larger than for the {\sl  AW} algorithm because they depend on the 
actual state which can take $q$ values.

We treat thereafter the following discrete spins: Ising $\pm1$,
Blume-Capel model, Potts model and clock model, and compare the 
efficiency of each algorithm.

\subsection{Ising {\boldmath $\pm 1$} spins}
We consider here the standard Ising spin system.
The spins $S_i=\pm1$ are under the influence of a local
field $h$ generated by the neighboring spins. They have an
energy
\begin{equation}
\label{E_Ising}
E_{i}\, = -\,T\cdot h \cdot S_i
\end{equation}
with the temperature $T$ and the local field
\begin{equation}
\label{E_Ising2}
h\, = \,\frac{J}{T} \sum_{<j>}S_j 
\end{equation}
where the sum is over the neighboring spins.
The probability $e^{-E_i/T}$ for the orientation of the spin
$S_i$ can be written without normalization as
\begin{equation}
P(\pm 1) \, = \, {\rm e}^{\mp h} \ .
\end{equation}
Following the last section we discuss the 
various algorithms and their implementations for the Ising spins
accessible at our  homepage \cite{homepage}.

Since there are only two possible states the restricted heat bath
{\sl  DHB}$_\delta$ and the restricted Metropolis {\sl Me}$_\delta$ 
algorithms are the same. Moreover no Walker aliases are needed.
This is different for a multi--spin update for four spins on a square with
$2^4=16$ possible configurations. Each of the four spins has two
neighbor spins outside which contribute to $h$ the values
$-2/T$, $0$ or $2/T$. We implement for this case both the heat bath
{\sl AW}$_4$ and the restricted heat bath {\sl  AWH}$_{4\delta}$
algorithms using Walker method. 

We put the results of the simulations for a square lattice with a size
$L=10$ in fig.~\ref{Ising}.
The integrated autocorrelation times $\tau$ (see \cite{Loison_Simon} for its determination) are displayed in fig.~\ref{Ising}(a) for comparing the different algorithms. The
maximum at the critical temperature  $T_c$ is indicated by the black squares.
As to be expected $\tau_{Me}>\tau_{D\!H\!B}$ because the heat bath method
chooses the new state according to its probability. This property is
general and therefore  also $\tau_{D\!H\!B} > \tau_{D\!H\!B_\delta}$ because the new 
state is better chosen by the restricted method {\sl  DHB}$_{\delta}$ than
by the direct method {\sl  DHB}.

For Ising spins this is not difficult to analyze.
Imagine that the spin state is $-1$ and the local field 
$h$ positive. The probability to get to the $+1$ state is 
${\rm e}^{-h}\big/({\rm e}^h + {\rm e}^{-h}) =1\big/({\rm e}^{2h}+1)$ with the heat bath
{\sl  DHB} whereas simply $e^{-2h}$ for the restricted heat bath {\sl  DHB}$_\delta$. 
Since $e^{-2h}>\ 1\big/({\rm e^{2h}+1})$ an Ising spin is more often flipped
in the latter case. Similarly if
the actual state is +1, the probability to get the state -1 is
$1\big/({\rm e}^{-2h}+1)  <1$ with the heat bath {\sl  DHB} 
and strictly 1 with {\sl  DHB}$_\delta$ or the equivalent restricted
Metropolis procedure. These differences are small if $h \gg 1$ 
but become noticeable for smaller $h$. For the two-dimensional
ferromagnet on a square lattice small means already the critical temperature
where a third of the flips occurs for $h=0$.
This fact  explains also the difference in the acceptance
rate between {\sl  DHB} and {\sl  DHB}$_\delta$ 
shown in fig.~\ref{Ising}(c),

It is evident by comparing fig.~\ref{Ising}(a) with fig.~\ref{Ising}(c)
that a bigger acceptance rate corresponds to a smaller autocorrelation time.
One can also see that a two times smaller acceptance rate double
more than twice the autocorrelation time. The reason may be
that some parts of the lattice flip more than average,
some parts less, and these slow spins have a dominant influence on $\tau$. 
We will come back to this point when we study the continuous spins.

The multi--sites heat bath {\sl  AW}$_4$ and {\sl  AWH}$_{4\delta}$ improve
the situation compared to the single site {\sl  DHB}, but are somehow less efficient
than the {\sl DHB}$_\delta$ or the restricted Metropolis {\sl Me}$_\delta$ procedure.
This procedure, usually called simply ``Metropolis'' in the literature is 
therefore the most efficient algorithm for $\pm1$ Ising spins.

For the judgment of the numerical efficiency 
the important parameter is not the autocorrelation
time, but the actual time the computer consumes.
If an algorithm $\cal A$ has a
$\tau$ two times smaller than an algorithm $\cal B$, but is ten times slower
to execute, it is better to use the algorithm $\cal B$. 
We take as consumption time  
the product of the autocorrelation time $\tau$ by the actual 
simulation time of one Monte Carlo step.

The simulation time is plotted in fig.~\ref{Ising}(b).
For one MC step {\sl Me}, {\sl  DHB} and {\sl  DHB}$_\delta$ = {\sl Me}$_\delta$
algorithms use almost the same time. 
As to the results for {\sl AW}$_4$ and {\sl AWH}$_{4\delta}$
the four spins algorithm {\sl  AW}$_4$ is $40\%$ faster
than the three previous one spin algorithms
since the four spins are flipped simultaneously.
The benefit is lost with the {\sl  AWH}$_{4\delta}$ method even
if implementations are almost the same. The disadvantage
is due to the limited fast cache memory of the processor
with an access time 10 times faster as ordinary memory.
Using the Walker's method the informations to 
be stored becomes 16 times larger for {\sl  AWH}$_{4\delta}$ than
for the {\sl  AW}$_4$.
For the same reasons, when the temperature increases,
the number of configurations simulated increases, and the number of
variables stored in the cache memory overflows its capacity. 
This explains the behavior of {\sl  AWH}$_{4\delta}$.

Finally we show in fig.~\ref{Ising}(d) 
the consumption time of the computer 
to compare the efficiency of each algorithm.
At the critical temperature, the restricted heat bath {\sl  DHB}$_\delta$ 
and the restricted Metropolis $Me_\delta$ are 4 times more efficient 
than the Metropolis {\sl Me},
and $2.5$ times more efficient than the heat bath {\sl DHB}.
It is interesting to compare the efficiency of {\sl  AW}$_4$ and {\sl AWH}$_{4\delta}$.
The last algorithm has a smaller autocorrelation time than the former,
but a larger time of simulation and therefore it is less efficient.
Even the {\sl AW}$_4$ is not really more efficient than 
{\sl DHB}$_\delta$ = {\sl Me}$_\delta$. Therefore for the $\pm1$ Ising spin 
the restricted heat bath or Metropolis algorithm should be used since also
a single--site algorithm is simpler to program.

We checked that these results do not change markedly as a function of
the size. This is expected since all the algorithms we compare are
local, the autocorrelation times have the same behavior and their
comparative efficiency remains the same.

In addition we tested the two and three-dimensional
antiferromagnetic triangular lattices and the three-dimensional $\pm J$ 
spin glasses on a cubic lattice. The results are alike and
even more in favor of the {\sl DHB}$_\delta$ = {\sl Me}$_\delta$ compared
to {\sl  DHB} and {\sl Me}. This is so
because the number of zero local fields is more important 
in these cases, up to 30\% at low temperatures for the triangular
antiferromagnetic lattices. 

\subsection{General Ising spin and the Blume--Capel model}

\noindent The Blume--Capel model is defined by a Hamiltonian
\begin{equation}
H = -J \sum_{<ij>} S_i S_j + \Delta \sum_{<i >}S_i^2
\end{equation}
with the first sum over nearest neighbor pairs and 
the second  one over all $N$ spins.
Each spin has $2\,S\! + \!1$ components $S_i = -S,-S\! + \!1, \cdots\,S$. 
This model can be used to describe a mixture of He$_3$ and He$_4$.
Initially the model had three components  \cite{Blume}.
Later the model was extended from $S=1$ to $S=3/2$ \cite{Blume1.5}
and could be generalized to any $S$.

We apply the single--site algorithms defined previously.
For this model, contrary to the Ising model, the 
restricted Metropolis {\sl Me}$_\delta$ is no more identical to the restricted
heat bath {\sl  DHB}$_\delta$ and its corresponding Walker method {\sl  AWH}$_\delta$.

In fig.~\ref{Ising_S_a} we present results for this model.
The graphs for acceptance rate are shown in fig.~\ref{Ising_S_a}(a), equivalent to the
Ising case  in fig.\ref{Ising}(c). The simulations are for a square lattice of size $L=10$
for $S\! = \!-1,0,1$ with $\Delta \! = \! 0$. {\sl AWH}$_\delta$ and 
{\sl DHB}$_\delta$ have the same acceptance rate since they are only 
variants of the same restricted heat bath algorithm. This is also the case
for the heat bath algorithms {\sl DHB} and {\sl  AW}.
The  autocorrelation times $\tau$,
not shown here, follow a similar pattern as the acceptance
rates and therefore
$\tau_{AW\!H_\delta}=\tau_{D\!H\!B_\delta}<\tau_{AW} =
\tau_{D\!H\!B} < \tau_{Me_\delta} < \tau_{Me}$.

In fig.~\ref{Ising_S_a}(b)
the consumption time is shown for the same model, that is $S=\pm1,0$
and $\Delta=0$. At the critical temperature the restricted heat bath 
algorithms {\sl DHB}$_\delta$ and {\sl AWH}$_\delta$ have the same efficiency, 
whereas the heat bath algorithms {\sl  DHB} and {\sl  AW} are
$25\%$ less efficient and the
restricted Metropolis method $50\%$. 
The Metropolis algorithm, usually used in the literature
is 2.7 times slower.
The consumption time at the critical temperature if
$\Delta\neq 0$ in the fig.~\ref{Ising_S_a}(c)
exhibits similar features as seen in fig.~\ref{Ising_S_a}(b), which are
even more pronounced in favor of the restricted heat bath.

In fig.~\ref{Ising_S_a}(d) the consumption time at the critical temperature for $\Delta=0$
is displayed as function of $S$.
For very large $S$ the restricted and not restricted algorithms
give similar autocorrelation times. 
However, the  simulation times
for one step are not the same for the algorithms
and in particular the Walker algorithm {\sl  AW} and {\sl  AWH}$_\delta$ should 
give better results. For intermediate values of $S$ shown
we observe that all the heat bath algorithms for $S\geq 3/2$ are almost
equivalent and much more efficient that the restricted Metropolis {\sl Me}$_\delta$,
which is more than $1.5$ times slower, and the simple Metropolis {\sl Me}
more than a factor $2.3\,$. 

Since the algorithms are local these results hold for other lattices as well and
larger sizes. In the next section some results are shown as a function of the
size and for different lattices.

\subsection{Potts model}

\noindent The Potts model \cite{Potts} is defined by a Hamiltonian:
\begin{equation}
\label{eq_H_Potts}
H\, = -J \,\sum_{(ij)} \delta_{q_i \,q_j} 
\end{equation}
$\delta_{q_iq_j}$  refers to the $q$ state Potts spin with
$\delta_{q_iq_j}=0$ when $q_i\ne q_j$ and 
$\delta_{q_iq_j}=1$ when $q_i = q_j$. 
For the two-dimensional square lattice with ferromagnetic interactions
the transition is of second order for $q\leq 4$ and of first 
order if $q>4$. This model is still extensively studied in 
particular with antiferromagnetic interactions \cite{Potts_today}. 

In fig.~\ref{Potts_a} we plotted the result for the ferromagnetic three state
Potts model ($q=3$) on a  square lattice (2s).
The system size is $L=20$. 
The results are very similar to the Ising two states case, but 
as clearly seen in fig.~\ref{Potts_a}(b)
the Walker method improves the situation 
compared to the heat bath by a gain
of $30\%$ for the direct heat bath and of
$50\%$ for the restricted heat bath. From the fig.~\ref{Potts_a}(d), 
at the critical temperature the restricted heat bath
with Walker's alias {\sl AWH}$_\delta$ is $50\%$ more
efficient than the same heat bath without Walker's alias {\sl  DHB}$_\delta$,
and more than two times better than the heat bath procedures {\sl  AW} and {\sl  DHB}.
The factor increases to more than three for the restricted Metropolis {\sl Me}$_\delta$ and to six
when Metropolis $Me$ is used. Obviously this ratio depends on the
implementations and the ones used here are accessible at \cite{homepage}.

For the same model and at the critical temperature  in fig.~\ref{Potts_b}(a)
the graphs for the consumption time are shown as a function of the system size $L$. 
One observes that the results do not depend on the size in a marked way.

In fig.~\ref{Potts_b}(b) we show the consumption time at the critical
temperature as a function of $q$. For $q=10$ the system exhibits a
strong first order transition and we have used 
the multicanonical algorithm \cite{Berg} in combination
to the local algorithms.
For large $q$ restricted algorithms are equivalent to non
restricted ones. The heat bath algorithms
are much better, $\approx 8$ times faster for
$q=10$, than the Metropolis ones. We observe an increase
of the consumption time for {\sl  AWH}$_\delta$ algorithm for $q=10$.
This is due to the
size of arrays needed to store which exceeds the possibilities of the
cache memory. For more details see the Ising $\pm1$ section and the
problems for the {\sl  AWH}$_{4\delta}$.

Since all the algorithms are locals the results hold for other
lattices and interactions as well. 
In the fig.~\ref{Potts_b}(c) and fig.~\ref{Potts_b}(d) we plotted the
results for the acceptance rate and the consumption times for the
anti--ferromagnetic three state Potts model on a  three-dimensional
cubic lattice (3ca). 
We get similar results at the critical temperature with a gain of four when we compare the {\sl  AWH}$_\delta$ to the Metropolis algorithm {\sl Me}.

\subsection{Clock model \label{clock_model}}

The $q$ state clock model \cite{LandauClock,Clock}
is a discrete version of the continuous $XY$
model. The Hamiltonian may be written as
\begin{equation}
H = -J\sum_{<ij>} \cos[2\pi(q_i-q_j)/q\,] \ .
\end{equation}
For $q=2$ this is the Ising model and for $q\rightarrow\infty$
the $XY$ model.
An example for $q=3$ is shown in fig.~\ref{Clock_a}(a).

We adapt  as before different local algorithms to this model.
In fig.~\ref{Clock_a}(b) we plot an example for the Walker
``boxes''  described in the section ref~\ref{section_Walker}.
For this model it is advisable to change the restricted
Metropolis method in the sense that a new state $q^{new}$ is confined
to the neighborhood of the old state $q^{actual}$. The simplest way is
to take  the new position within the limits 
$q^{actual} -\delta \leq q^{new}\leq q^{actual} +\delta$ with
$\delta$ varying from $1$ to $q/2$.

In fig.~\ref{Clock_a}(c) the autocorrelation time is displayed for the
10 state clock model with ferromagnetic interactions on a
square lattice (2s) as a function of the temperature $T \propto 1/h$ and $\delta$. 
At low temperatures the best algorithm is the one
which tests only new states close to  the actual one and at high temperatures
it is better to test all possible states. At  temperatures between
an intermediate $\delta$ works best.

Before looking at  fig.~\ref{Clock_a}(d) we turn to
fig.~\ref{Clock_b}. The model tested there is the six state 
clock model on the two-dimensional square lattice (2s) of size $L=10$
with ferromagnetic interactions. It is clear that the Walker method improves
considerably the heat bath methods as can be seen in fig.~\ref{Clock_b}(b).
Obviously this depends on the implementations and the ones
used in these figures
are accessible at our homepage \cite{homepage}. The gains at the critical temperatures
(there are two critical temperatures  \cite{LandauClock}) are quite
appreciable in using the heat bath Walker method {\sl AW}. This method is at least
four times faster than the Metropolis methods as can be seen in
fig.~\ref{Clock_b}(d). 
In the figures the {\ sl Me}$_{\delta*}$ corresponds to the optimal
choice of $\delta$  for each temperature.

Now we come back to the fig.~\ref{Clock_a}(d). The consumption times
at the critical temperature (the higher temperature of the two
critical temperatures) are displayed as function of the number of
states. Whatever $q$ is, we observe that the Walker heat bath methods are 
much more efficient than any other method. Moreover for large $q$ we
notice that the restricted Walker heat bath {\sl  AWH}$_\delta$ algorithm
becomes less efficient than for low $q$. The reason is that
the arrays become too big to be stored in total in the cache memory,
a problem we encountered before (see the Ising $\pm$ section).
Interesting is also the result for the $q=2$ case corresponding to the Ising
model. We should get identical results for {\sl  AWH}$_\delta$ and {\sl Me}$_\delta$
since they are the same algorithm in this case. Indeed the same
autocorrelation times are  found but the simulation times differ and
therefore also the consumption times. The reason is that our implementation which
treats any $q$ does not use special features of the $q=2$ case. 
Again the reader should be aware that a good implementation is as
important as a good algorithm.

\section{Continuous spins}

\noindent We study now continuous spin systems. The Hamiltonian $h$ should have the
form
\begin{equation}
\label{good_form1}
H\, = -\, T\,\sum_i \hbox{\boldmath $h$}_i\cdot \hbox{\boldmath $S$}_i 
\end{equation}
similar to (\ref{E_Ising}) and (\ref{E_Ising2}) with 
\begin{equation}
\label{good_form2}
\hbox{\boldmath $h$}_i = \, \frac{1}{T} \sum_j  J_{ij} \cdot \hbox{\boldmath $S$}_j \ .
\end{equation}
A magnetic field term or any other potential function
depending only on norm of the spin $S_i$ can be added.
We note that it is possible  to write the dipolar spin spin interaction this way.
This form is dictated by most of the improved simulation methods we are
going to discuss.
For more details about this limitation we refer to the 
section \ref{other_type_H}.

\noindent The probability to find the spin $i$
in the state {\boldmath $S$}$_i$ is
given by the Boltzmann factor
\begin{eqnarray} 
P(\hbox{\boldmath $S$}_i) \cdot d \hbox{\boldmath $S$}_i = \,
\hbox to 2.2cm{${\rm e}^{\,\hbox{\boldmath$\scriptstyle h$}_i \cdot 
\hbox{\boldmath$\scriptstyle S$}_i - h_i} \ \cdot\, d$ {\boldmath $S$}$_i$} &\ &\nonumber \\
\label{P_O1}
             = \,\hbox to 2.2cm{${\rm e}^{\,h\cdot x - h}\ \cdot dx$\hfill} 
			&\ &\text{for continuous Ising spins $x \in [-1,1]\,$,} \\
\label{P_On}
	   = \,\hbox to 2.2cm{${\rm e}^{\,h\cdot \cos\,x-h}\ \cdot\, d\,\Omega$\hfill}  
			 &\ &\text{for $O(N)$ spins with $N \geq 2\,$}
\end{eqnarray}
with 
\begin{eqnarray} 
\label{Omega_O2}
d\Omega \, = \,\hbox to 3.2cm{$dx$\hfill }
            &\ &\text{ for $N = 2$ with $x \in [-\pi,+\pi]$,} \\
\label{Omega_O3}
           = \,\hbox to 3.2cm{$\sin x\, dx \cdot dy$\hfill }
           &\ &\text{ for $N = 3$ with $x \in [0,\,\pi]$ \& $y \in [0,\,2 \pi]$,} \\
\label{Omega_O4}
          = \,\hbox to 3.2cm{$\sin^2\! x\,dx \cdot \sin y\,dy\cdot dz$ \hfill }
          &\ &\text{ for $N=4$ with $x,y \in [0,\,\pi]$ \& $z \in [0,\,2 \pi]\,$}
\end{eqnarray}
and similarly for $N>4$. In eq.~(\ref{P_O1}) and (\ref{P_On}) the suffix $i$ 
has been dropped. 
The additional factor ${\rm e}^{-h}$ in the definition of the probability $P$
is a kind of normalization.

\subsection{Simulation methods}
In this section we will present the methods to be tested on continuous spin systems. 
The starting point are again eqs.~(\ref{detailed_balance}--\ref{Hasting}). 

\subsubsection{Metropolis algorithm} As for discrete spins, the
easiest algorithm to implement is the Metropolis method {\sl Me}. It
consists of choosing randomly new spin directions and applying the
condition (\ref{Hasting}).  For $N\leq 3$ this is not a difficult task.
However for $N>3$ we cannot integrate and invert the weight of
$d\Omega$ in Eq.~\ref{Omega_O4}. In this case the standard
procedure is to calculate the new spin direction using Gaussian random
numbers.  More details about the different spin types will be found in the
corresponding section. For any system the Metropolis procedure 
is much less efficient, at least by a factor two, than the algorithms
we will discuss below.

\subsubsection{Restricted Metropolis algorithm}

The restricted Metropolis algorithm {\sl Me}$_\delta$ is here defined
in a similar way as for the clock model in section~\ref{clock_model}.
The new state is chosen around the actual one
restricted by
$x^{actual}-\delta \leq x^{new}\leq x^{actual}+ \delta$
using Hasting's method
(\ref{Hasting}). This method is usually applied for the $XY$
spin $N=2$, but could be used for any $N > 2$.
The results obtained are better than for 
the Metropolis algorithm but still this method is
less efficient than the ones
introduced further down.

\subsubsection{Direct Heat Bath}
The direct heat bath {\sl  DHB} can be
used only for continuous Ising spins $-1\leq S <1$ and
Heisenberg spins with 3 components putting simply  $f(B) = P(B)$ in
(\ref{heat-bath}). 
The probability is for both cases
 eq.~\ref{P_O1}, that is $P(x)\cdot dx={\rm e}^{h\cdot
x-h}\cdot dx$ with $x\in[-1,1[$. The cumulative probability
$F$ given by the normalized integral of $P$ is then
\begin{eqnarray}
\label{cumulative}
F(x)&=&\frac{\int_{-1}^x P(\tilde x)\,d\tilde x}
            {\int_{-1}^{+1} P(\tilde x)\,d\tilde x} \nonumber \\
	\label{F_O1}
	&=&\frac{{\rm e}^{h\cdot x}-{\rm e}^{-h}}
	        {{\rm e}^{+h}      -{\rm e}^{-h}}
\end{eqnarray}
so that $0 \leq F(x) < 1$.
In fig.~\ref{O1_a}(b) $F(x)$ is plotted.
From a random number $F= \ran$ between 0 and 1 one can then determine
$-1 \leq x < 1$ by inverting (\ref{F_O1}):
\begin{eqnarray}
 \label{x_O1}
x &=& 1+\frac{1}{h}\cdot\log((1-{\rm e}^{-2h})\cdot \ran+{\rm e}^{-2h}) ) \ .
\end{eqnarray}
This formula has been used for an implementation \cite{Miyatake}
which actually is easier than for the discrete spins described in the
previous section \ref{DHB_discrete}.
With a slight modification one can improve the efficiency especially for
large $h$ or low temperature. One generates a value for $x$ in
the limit $-\infty < x < 1$ by
\begin{eqnarray}
\label{x_O1_best}
  x=1+\frac{1}{h}\cdot\log(\ran) 
\end{eqnarray}
and exclude the values outside the range [-1,1].
This procedure must be iterated until one obtains a value between
$-1\leq x < 1$.
This way one does not need to calculate ${\rm e}^{-2h}$ but a
number of choices are rejected and additional random numbers must be
generated.
One assumes here that $h$ is positive so that the probability
$P(x) \propto {\rm e}^{\,x\cdot h}$ can be extended to arbitrary 
negative values of $x$.
If $h$ is negative one has to change the sign and put
$x \Rightarrow x \cdot {\rm sign}(h)$. At the critical temperature and 
below typically
one has $h > 1.5$ for various models and the efficiency increases  by 
25\% using the simpler rule (\ref{x_O1_best}) instead of (\ref{x_O1}).

For vector spin models with $n = 2,\, 4,\,$ and more components the 
direct heat bath method
cannot be applied since the weights are too complicated for a simple 
integration. One is
therefore forced to choose another form for the function $f$ in 
Eq.~\ref{heat-bath}
and use the
rejection method. The choice is very broad and many functions have been 
tried especially
for $XY$--spins with two components. However, the function chosen are 
particular for each model.
We introduce below three completely general algorithms applicable to 
any model with
an energy of the form given in (\ref{good_form1}-\ref{good_form2}).

\subsubsection{Fast linear algorithm}

For an efficient heat bath simulation a function $f>P$ should be chosen
for which the integral is easy to determine and also the inversion poses
no problem. For a maximum gain in simulation time we find that the best
choice is a function with $n$ steps.  For an example see
fig.~\ref{O1_a}(a) and (c) where we present choices of step functions
for the continuous Ising spin $S$ ($-1\leq S <1$) and $h=5$.  
The idea is to use this together with 
the Walker algorithm or a variant of it.
We first describe the method, called the Fast Linear Algorithm
({\sl FLA}), in detail.

\paragraph{$h$ is constant}
We first consider the case $h={\rm const}$ where $h$ is defined in
eq.~\ref{good_form2}, and then show later  how to deal with the
real situation where $h$ can vary.

The function $f$ is made of $n$ constant
parts $f_i$. We choose the $f_i$ with two conditions:
\begin{eqnarray}
\label{equ_f_i_max}
&&f_i = \text{ maximum of } q(x) \text{ in the interval } [x_i,x_{i+1}] \, ,\\
\label{equ_f_i_cst}
&&f_i\cdot (x_{i+1}-x_i)= a \, ,
\end{eqnarray}
with the constant area $a$ to be determined. 
For an example see the fig.~\ref{O1_a}(a).
The first equation ensures that $f$ is always greater
than $q$. The second is useful for a simple and faster way to invert
the cumulative probability $F$.
Indeed  we see that $F$ is composed of $n$
intervals of straight lines (integration of $n$ constant functions). 
With (\ref{equ_f_i_cst}), all intervals are equal as
shown in fig.~\ref{O1_a}(b).
Therefore, only a few steps are necessary to invert 
$F$. This is done as follows.\\
Choosing a random number 
between 0 and 1, we find the interval corresponding $i$ 
\begin{eqnarray}
\label{i_int_F}
i={\rm int} (n\cdot \ran)
\end{eqnarray}
where {\rm int} means the conversion from real to integer. Having $i$, 
it is not difficult to find $x$ that we are looking for:
\begin{eqnarray}
\label{x_a_b}
x&=&x_i+(n\cdot \ran - i)\cdot (x_{i+1}-x_i) \ .
\end{eqnarray}

Since we use only conversions and simple operations, this method is  
extremely fast. We come back to this point later when 
we conjecture that it is impossible to find a faster 
algorithm for this class of problems.
Moreover, for $n \rightarrow \infty$ the acceptance rate, 
i.e., the area of $P$ divided by the area of $f$ 
between $[-1:1[$, tends to $100\%$. Typically with 
$n$ being equal to some hundreds one obtains a rejection rate 
($=1-$ acceptance rate) of few percents.

Technically, we find the $\{x_i\}$ and the $\{f_i\}$
using (\ref{equ_f_i_max}-\ref{equ_f_i_cst}) and  
store the value of $a$ and $\{x_i\}$.
This must be done once before the Monte Carlo 
simulation begins.
Then apply (\ref{i_int_F}) and (\ref{x_a_b}) at each Monte Carlo step,
calculating $f_i$ using (\ref{equ_f_i_cst}).

To determine the $\{x_i\}$ we use an iterative procedure. 
First we fix $a$. Next, using (\ref{equ_f_i_max}) and (\ref{equ_f_i_cst}),
we calculate the corresponding $\{x_i\}$ and $\{f_i\}$.
When $x[n]>1$ we reduce $a$, otherwise we increase it.
The procedure is stopped when $1 \leq x[n] \leq 1+\epsilon$, with
$\epsilon$ being the accepted error.
Then, we fix $x[n]=1$ and $f_{n-1}=\frac{a}{1-x[n-1]}$. 
Since $\epsilon$
is positive, $f_{n-1}$ obeys (\ref{equ_f_i_max}).
One possibility is to put the initial value of $a$ at $a=A/n$, $A$ being 
the area of $P(x)$ between $[-1,1]$ calculated numerically.

\paragraph{$h$ is variable}
Now we have to consider the case where $h$ is no longer constant. 
We notice $P_{h_1} > P_{h_2}$ if $h_1<h_2$ with $h_i$ 
the norm of the vector ${\bf h_i}$, therefore $h_i\geq 0$. 
We divide the possible range of variation of $h$ 
in $M$ parts corresponding 
to $h_1$, $h_2$, \ldots,$h_M$ and use the table calculated at $h_j$
for $h_j \leq h < h_{j+1}$. Due to this procedure
the acceptance rate decreases. 
We decide to choose the ($\{h_j\}$) 
so that the rejection rate is less than a threshold using:
\begin{eqnarray}
\frac{\text{area of } P_{h_j}-\text{ area of } 
P_{h_{j+1}}}{\text{area of } P_{h_{j}}}= {\rm threshold}\, .
\end{eqnarray}
The area of $P_h$ is calculated numerically.

However, the values  $\{h_j\}$ are not linearly distributed. 
Therefore, we create a new table ${\rm table_h}$ with $K$ elements. 
In each Monte Carlo run we calculate the value $k$ with a conversion
from real $h$ to integer $k$
and the ${\rm table_h}$ gives us the value of $h_i$.
More specifically: 
\begin{eqnarray}
\label{find_k}
k&=&{\rm int}(K\cdot (h-h_0)) \, ,\\
\label{find_h}
h_i&=&{\rm table_h}[k]\, ,
\end{eqnarray}
where $h_0$ is the first value of $h_j$. The ${\rm table_h}$ has to be filled 
before the beginning of the Monte Carlo simulations.
$K$, the number of elements 
of the table, is free, but must be large enough to be able to
differentiate the $h_j$. 

In summary, we have a way to generate the value $x$ with a probability 
$f(x)$ using (\ref{find_k}), then
(\ref{find_h}), then (\ref{i_int_F}) and finally (\ref{x_a_b}).
Using this value of $x$ we apply the rejection method, 
i.e. accept the new value if $\frac{P(x)}{f(x)}\geq \ran_2$, where $\ran_2$ is
a random number in the interval $[0,1[$.
We call this algorithm the Fast Linear Algorithm because
we use only linear equations.
Few steps are necessary to obtain the final value $x$
looking at the C program given at our homepage
\cite{homepage}.

\paragraph{Advantages and flaws}

First we would like to argue that no other algorithm 
using the rejection method could be faster than 
the Fast Linear Algorithm.

The rejection method needs two 
uniformly distributed random numbers at each Monte Carlo step. 
Step  1 is to get a value $x$ from a known cumulative 
of the test function $f$ and
step 2 is to compare the value $f(x)$ to the probability.
To compare different algorithms using this method, it 
is enough to calculate the time necessary to perform step 1
and step 2 without considering the time to produce the random 
numbers since it will be the same for all algorithms. 

The consumption time can estimated in comparing 
the various possible operations executed by the computer.
We take the following equivalences: basic operation
(\verb#*#, \verb#+#, \dots) = 1\thinspace unit,
conversion from real to integer (\verb#int#) = 3\thinspace units,
(\verb#if#) = 3\thinspace units, 
square root (\verb#sqrt#) = 6\thinspace units, and calculating a function 
like cosine, exponential or logarithm = 13--18\thinspace units. These estimates 
are only approximate, but sufficient for comparing
the different algorithms. 

To apply the {\sl FLA} one needs two conversions from real to integer
according to eq.~(\ref{i_int_F}) and (\ref{find_k}),
and a few basic operations like additions, multiplications,
and the use of tables. 
If we count the time necessary
for our algorithm very roughly, we get 20 ``units''. 
Therefore, the time necessary for one step of the {\sl FLA}--algorithm
is comparable to the time needed to calculate one function. 
Another algorithm, to be as fast as {\sl FLA}, 
should use only once a calculation of a function. 
However, in most of the cases, it will never reach such a
high rate of acceptance comparable to 
{\sl FLA} of nearly $100\%$.
Therefore the algorithm {\sl FLA} should be always the fastest 
algorithm. This conclusion holds only when the rejection method is used, 
not for the heat bath methods when the probability can be integrated and
inverted as for a classical Heisenberg model. However, even in this case
it will be shown in the following that this heat bath algorithm will not be faster
than the {\sl FLA} algorithm.

We note another advantage of this method. It is completely 
general and can be applied to all types of probabilities.
Moreover, we have introduced the {\sl FLA} with a probability
on a fixed range $[-1,1[$.
However, it is possible to handle also the case of an infinite range with a 
somehow more complex program, as can be seen in the section dealing with
the Ising $\phi^4$ model.

The only flaw of {\sl FLA} is that a certain amount
of memory must be available to store the tables.
For a Hamiltonian which can be factorized like in
eq.~(\ref{good_form1}--\ref{good_form2}) there are two variables, $x$ and $h$,
to handle. 
Typically we get files which are less than 
100 Kb for a rejection rate of about $15\%$ percents.
If one wants to decrease this rate 
the size of the tables increases 
and the cache memory could become too small to store the arrays.
A test of 
the maximum number of steps which minimize the consumption time 
is needed at each simulation. See for an
example fig.~\ref{O1_b}(b).

\subsubsection{Walker's Algorithm}

The last algorithm is very efficient, but it is not so easy to
calculate the $\{x_i\}$ and it becomes really complex in presence of more
variables ${x_i,y_i,\cdots}$. In this case the Walker algorithm could
provide a good solution. 
The only difference between the Fast Linear Algorithm ({\sl FLA}) and the
Walker algorithm ({\sl  AW}) is the choice of ${x_i}$. Instead of
(\ref{equ_f_i_cst}) we choose the ${x_i}$ at fixed interval, for example
on the fig.~\ref{O1_a}(c) the dashed line (``{\sl  AW} without $x_H$''). 
Another possible choice to avoid to get a lot of informations in the
region where the probability is almost zero 
is to choose $x_1=x_H$ and the others $x$
at fixed intervals  between $x_1$ and 1 (''{\sl  AW} with $x_H$''). 

Whatever the choice is, since (\ref{equ_f_i_cst}) is no more
valid, we cannot use only (\ref{i_int_F}) 
to calculate $i$. We must use the
Walker's alias introduced previously (section \ref{section_Walker}). 
We must calculate the integral of each step function, i.e. the area
$f_i\cdot(x_{i+1}-x_i)$. Walker's algorithm is used to divide
them in $n$ equal boxes as shown in fig.~\ref{O1_a}(d) for the step 
function ``{\sl  AW} with $x_H$'' shown in fig.~\ref{O1_a}(b). 
During the simulation, we
have to use, as for the discrete spins, an \verb#if# condition to 
choose the correct ``state''. Therefore in addition of (\ref{i_int_F})
we must use:
\begin{eqnarray}
\label{i_int_F_Walker}
{\rm if}(\ran > P^i_{limit}) \,\,\,\,  i={\rm Alias}[i] \, .
\end{eqnarray}
And (\ref{x_a_b}) must replaced by
\begin{eqnarray}
x&=& x_i^{box,ini} + \ran\cdot(x_i^{box,fin}-x_i^{box,ini})
\end{eqnarray}
with $x_i^{box,ini}$ and $x_i^{box,fin}$
are the limit for each box and ``state''.
Otherwise the algorithm is similar to the {\sl FLA}.

The difference of this algorithm and 
the {\sl FLA} is that we have to store much more data in
the arrays. Precisely we have to save the $\{P^i_{limit}\}$,  
$\{x_i^{box,ini}\}$, $\{x_i^{box,fin}\}$,
\{Alias$[i]\}$ and also $\{f_i\}$ to apply the
rejection method. This must be compared to  $a$ and $\{x_i\}$ for
the {\sl FLA}. 
As a consequence the arrays are $6$ times bigger for the {\sl AW} 
than for the {\sl FLA}. 
To be efficient arrays must be stored in the cache
memory and therefore the {\sl AW} is less efficient than the {\sl FLA}. 
For example see fig.~\ref{O1_b}(d) and fig.~\ref{O2_b}(b).

However the arrays are very easy to create contrary to the ones for the 
{\sl FLA}, even with more than one variable.
We will present two examples with two
variables below 
when treating Heisenberg spins with a $\phi^4$ potential.

\subsubsection{Walker's Algorithm with Hasting's method}
Another advantage of the Walker algorithm over the {\sl FLA} is
that we can
use Hasting's method. Indeed to apply this last method we need to be
able to get not only $F^{-1}(\ran)$ but also $f(x^{actual})$
as seen in the Eq.~\ref{Hasting}.
To obtain easily $f(x^{actual})$
we need a simple law for $x_i$, the simplest
being the $\{x_i\}$ distributed at constant intervals.
Then we use the Hasting's method (\ref{Hasting})
in combination with the Alias Walker.
We  call it, as
before, the Alias Walker Hasting ({\sl AWH}) algorithm.

One big advantage of the Hasting's method compared to the heat
bath-rejection method is that we do not need to choose the function $f$
bigger than the probability $P$. This proves very useful if the
Hamiltonian cannot be factorized like
$H={\bf h}_i\cdot {\bf S}_i$ (\ref{good_form1}).  
In this case the {\sl FLA} cannot be
applied but the present method works.

In the next section we present the vector spin with dimensions $N$
from 1 to 6 separately.
We try to be as complete as possible in order that the 
reader can use the proposed methods easily. For each case
the C programs are accessible at our homepage  \cite{homepage}.

\subsection{Continuous Ising spins {\boldmath $-1\leq S < +1$, $N=1$}}
The continuous Ising spin varies from -1 to +1 with an uniform 
probability so  that the Boltzmann weight according to (\ref{P_O1}) is
\begin{equation}
\label{P_O1_2}
P(S_i) \cdot dS_i  = {\rm e}^{h\cdot x-h}\cdot dx 
			   \text{ ,  } x\in(-1,1) \ .
\end{equation}

We explain  in more detail the results for this first example of continuous
spins since the vector spin cases $N>1$ discussed later will be technically
similar. We restrict ourselves to three types of algorithms.

The first is the Metropolis algorithm {\sl Me} where one  
chooses the new value $x^{new}$ uniformly between -1 and 1. 
This value is accepted with a probability $P(x^{new})/P(x^{actual})$ 
according to eq.(\ref{Hasting}) otherwise the old value $x^{actual}$
is kept. To gain a better understanding how this procedure works
we turn to fig.\thinspace\ref{O1_a}(a) where the probabilities for a
local field $h=5$ and $h=10$ are plotted. 
The Metropolis choice for the function
$f_{Me}=$``1'' is also shown.
Since Hasting's method is used, the 
function $f$ is automatically renormalized to $P(x^{actual})$.
If $P(x^{new})\geq P(x^{actual})$ the new state is always accepted,
which corresponds to $x^{new}\geq x^{actual}$. 
Otherwise  for $x^{new}< x^{actual}$ the new state is taken
with a probability $P(x^{new})/P(x^{actual})$. Graphically the
acceptance rate is equal to the ratio of the areas under 
$P(x^{new})$  and under the straight line $f_{Me}=P(x^{actual})$ 
for $f>P$.
If the local field $h$ increases or the
temperature decreases, this ratio decreases 
tending to zero for low temperatures as can be seen in the
fig.\thinspace\ref{O1_a}.

The second method, the heat bath {\sl DHB}, does not have this defect.
Formula (\ref{x_O1}) is used to find the new value $x^{new}$.
Here the rate of acceptance is $1$. Updating with
formula (\ref{x_O1_best}) instead creates difficulties since
small local fields  occur too frequently.
The cumulative probability $F$ is displayed in fig.\thinspace\ref{O1_a}(b).

The third type of simulation is Walker's method including the 
Fast Linear Algorithm {\sl FLA}, the Alias Walker algorithm {\sl AW} 
and the Alias Walker Hasting algorithm {\sl AWH}. For the
two last algorithms we choose $x_1$ with $P(x_1)=0.01$ 
and the others $\{x_i\}$ are fixed at constant intervals between $x_1$ and $1$. 
We plot such a construction in fig.\thinspace\ref{O1_a}(c) under the label ``{\sl AW}
with $x_H$''. This somewhat arbitrary choice of $x_1$ is recommended since
for $-1< x<x_1$ the probability is small enough and therefore it is not necessary 
to provide detailed informations on this range.
The corresponding Walker aliases are plotted in fig.\thinspace\ref{O1_a}(d).

The {\sl FLA} fixes automatically this last problem.
In fig.\thinspace\ref{O1_a}(a) the step
function $f_{FLA}$ is shown for five bins and
the corresponding cumulative probability in fig.\thinspace\ref{O1_a}(b) .
There the dashed lines are equidistant because of the condition (\ref{equ_f_i_cst}).
In this method the acceptance rate is the ratio of the area under $P$ and
under $f_{FLA}$ in fig.\thinspace\ref{O1_a}(a).
This ratio can be made close to 1 by increasing the number of bins. The
deviation from this ideal value we will denote by ``error'' in the following.

We compare now in fig.\thinspace\ref{O1_b} the efficiency of the different
algorithms for an antiferromagnet on a three-dimensional
stacked triangular lattice (3t).
A system size of $L=6$ is sufficient for the test.
Fig.\thinspace\ref{O1_b}(a) shows the
refusal rate of  {\sl FLA}, {\sl AW} and {\sl AWH} as a
function of the ``error'' given as input to the programs \verb#create_O1.out#
and \verb#create_O1_walker.out# \cite{homepage} used to create the arrays.
The actual refusal rate is approximately only half of the ``error'' value for the
{\sl FLA} and the {\sl AW}, and one fourth for the {\sl AWH}. 
It is worth to note that {\sl AW} and {\sl AWH} use the same information stored
in an array. The difference is that the Hasting algorithm is automatically renormalized
to the value $P(x^{actual})$ as explained before, and therefore it
fits the probability distribution $P$ better.

The fig.\thinspace\ref{O1_b}(b) shows the consumption time for the last
three algorithms  at $T = T_c \approx 1.2$.
The three curves have a similar behavior. At first, if the error 
decreases by increasing the number of bins 
the refusal rate will decrease, and therefore
the consumption time decreases also. 
However if the number of bins becomes too large, the
array cannot be kept in the fast cache memory, the time of simulation
increases and the gain in the acceptance rate does not compensate
the loss in the simulation time of one MC step. We remind the
reader that the consumption time is defined as the product of the
simulation time of one MC step and the autocorrelation time.
This last quantity being
shorter for a higher acceptance rate.
We observe that the {\sl FLA} has the smallest consumption time because 
the array to be stored is six times smaller. 

We display in fig.\thinspace\ref{O1_b}(c) also the dependence of the acceptance rate
for the various algorithm on the temperature or more precisely on the
inverse of the local field $1/h \propto T$.
The number of bins for the algorithms ({\sl FLA}, {\sl AW} and {\sl AWH}) is
chosen to minimize the consumption time.
The acceptance rate of {\sl Me} goes to zero when the temperature decreases as
has been discussed before. As a consequence the autocorrelation time
(not shown) increases without limit.
As can be seen in fig.\thinspace\ref{O1_b}(c) the {\sl AWH} acceptance 
rate is
larger than the {\sl AW} rate, even if  the same array is used.
The {\sl FLA} and the {\sl AWH} have a high acceptance rate for
any $h$ comparable to the acceptance rate of the direct heat bath {\sl DHB} 
which is strictly $1$.

There is a difference between the acceptance rate of the Hasting
algorithms {\sl AWH} and {\sl Me} on one hand,
and the heat bath algorithms {\sl AW} and {\sl FLA} on the other hand.
In the first case, after a refusal the same spin is kept. 
In the last case a new spin is chosen until it is accepted,
and therefore the new state is always different from the old one.
We will see the implication of this fact in the next paragraph.

The most important parameter in numerical simulations, the consumption
time,
is displayed in fig.\thinspace\ref{O1_b}(d). Clearly the {\sl FLA} algorithm is the most
efficient one even compared to {\sl DHB}. 
It it interesting to compare the result for {\sl Me} with the inverse of the
acceptance rate multiplied by the ratio of the times of simulation for
{\sl Me} and {\sl FLA}, plotted with the symbol \%. 
If the system were composed of only one spin, the two 
lines would collapse. We observe that the consumption time for $Me$ 
is above the \% line. Indeed, if the spins of some parts of the lattice
are flipping more than the average, some group of spins
flip much less than
the average and these last spins have a strong influence and the
autocorrelation time $\tau$, and therefore on the consumption time
for the simulation. 

We have tested different lattice sizes $L$ 
in order to check that results discussed do indeed not
depend on $L$, since the algorithms are local ones.
For different lattices the only change is the
value of the local field $h$. 
Looking at  our homepage \cite{homepage}
one can see the nearly collapsing curves 
for an anti--ferromagnet on a  two-dimensional triangular lattice (2t)
and a ferromagnet on a square lattice (2s).

In the literature, studies on the continuous Ising model are not very common.
To our knowledge the most recent study is 
for two-dimensional ferromagnetic interactions
of long range using the Metropolis algorithm \cite{Bayong}.
The use of  the heat bath methods {\sl FLA} or {\sl DHB} would increase
the efficiency  because an acceptance rate close to 100\% is even more
important if the interaction is of long range.
Since the simulations for the continuous Ising model ($N=1$) 
is similar to the Heisenberg case ($N=3$)
all conclusions for the former model hold also for the latter.

\subsection{Ising {\boldmath $\phi^4$} model, {\boldmath $N=1$}}
In this section we study the Ising $\phi^4$ model \cite{Hasenbusch}
or Ginzburg--Landau model on a lattice
where the spin variable $x$  varies from $-\infty$ to $\infty$.
The Boltzmann weight is 
\begin{equation}
\label{P_O1_Phi4}
P(S_i)\cdot dS_i = {\rm e}^{h\cdot x-x^2-\lambda (x^2-1)^2}\cdot dx 
\end{equation}
and in fig.\thinspace\ref{O1_Phi4}(a) this probability is plotted for $h=1$
and $\lambda=1.3182$.
It is interesting to study this model for its own sake.
Here it serves as an example how to cope with the infinite boundaries of
the probability distribution in the {\sl FLA}, {\sl AW} and {\sl AWH} algorithms.
Cluster algorithms exist for this model but one needs a
local algorithm to vary the norm of the spin \cite{Hasenbusch}. 
The probability (\ref{P_O1_Phi4}) cannot be integrated in a simple way
and therefore no direct heat bath method is available.

With the restricted Metropolis {\sl Me}$_\delta$
algorithm the new spin $x^{new}$ variable is chosen around
the actual one:
$x^{actual}-\delta \leq x^{new}\leq x^{actual}+ \delta$.
Then the Hasting formula (\ref{Hasting})
is applied with the function $f_{Me_\delta}=1$ plotted in
fig.\thinspace\ref{O1_Phi4}(a).
We compare here the efficiency of the restricted Metropolis algorithm
{\sl Me}$_\delta$ only with the Fast Linear Algorithm {\sl FLA} which is faster than 
{\sl AW} and {\sl AWH}.

The {\sl FLA} is slightly different from the one described 
in the previous section since a first step starting at $x_0=-\infty$
and a last one ending at $x_n=+\infty$ would lead to a divergence for the
cumulative function $F$. 
The solution is to take two 
exponential functions for the first interval $[x_0,x_1]$ and the last one
$[x_{n-1},x_{n}]$, and otherwise to take
a step function for $f$. An example is given in
fig.~\ref{O1_Phi4}(a) for $n=11$ steps. 
Unfortunately the implementation is also 
a little bit more complicated than previously.

Fig.\thinspace\ref{O1_Phi4}(b) gives the consumption time at different
temperature for the {\sl FLA} as function of the ``error'' given to
the program \verb#create_O1_Phi4.out# (accessible from our homepage
\cite{homepage}). The best choice is $20\%$ corresponding to $n=38$ bins.

The fig.\thinspace\ref{O1_Phi4}(c) and (d) display the acceptance rate and the
consumption time for a ferromagnet on a cubic lattice
(3c) for a system size $L=6$. We observe that the {\sl FLA} is much more
efficient than the Metropolis algorithm. At the critical
temperature represented by the squares, the Metropolis $Me_\delta$ is
almost three times slower than the {\sl FLA}.

Since our algorithms are local, results do not depend on the system
size nor on the type of the lattice. For this model there exist 
cluster algorithms and also an over-relaxation 
algorithm \cite{Overrelaxation}. We must employ them in combination (the
section devoted to $XY$ spins shows results for the {\sl FLA} in
combination with over-relaxation). Moreover if we are interested in
frustrated system there are no more cluster algorithms available and an
efficient local algorithm is fundamental.

\subsection{{\boldmath $XY$} and {\boldmath $U(1)$} variables, {\boldmath $N=2$}}
The $XY$ spin is a two-dimensional vector of norm one.
Since there is an equivalence between the direction of the spin and a
phase, both characterized by one angle, $XY$ spin systems and $U(1)$
gauge theory (see for example \cite{Creutz_Jacob}) have a common basis.
Due to the varied interests many algorithms have been tried.
We want to  compare them to 
the Fast Linear Algorithm {\sl FLA} which proves
to be the fastest algorithm, and the Walker Hasting algorithm
{\sl AWH}.

The Boltzmann probability is simply
\begin{equation}
\label{P_O2}
P({\bf S}_i)\cdot d{\bf S}_i  = {\rm e}^{h\cdot \cos x-h}\cdot dx  
\end{equation}
with the angle $x$ varying between $-\pi$ and $\pi$.

Contrary to the probability of the former
case $N=1$ given by (\ref{P_O1_2}) the cumulative probability
and its inversion is too costly in computer time and consequently
no direct heat bath method {\sl DHB} is possible.
One must use the heat bath rejection or Hasting's methods
instead.

We have tested first the standard Metropolis algorithms {\sl Me} for 
which $f(x)=1$ in eq.(\ref{Hasting}). The new angle $x^{new}$ is
chosen randomly between $-\pi$ and $\pi$ and accepted according to
the probability $P$ which is shown in fig.\thinspace\ref{O2_a}(a) for  $h = 4$.
If the local field $h$ increase, the probability is more peaked near the
origin and the acceptance rate decreases as explained before.
The restricted Metropolis {\sl Me}$_{\,\delta}$ gives better results.
In this case $x^{actual}-\delta \leq x^{new}\leq x^{actual}+ \delta$
as indicated in fig.\thinspace\ref{O2_a}(a).
For a small enough $\delta$  the acceptance rate can increases up to one,
but then the spin configuration will change very slowly
and the autocorrelation time becomes large. 
This situation is the same as for the clock model described in 
section~\ref{clock_model}.

For the heat bath method many different functions $f(x)>P(x)$ 
in the interval $(-\pi,\pi)$ have been proposed.
For the auxiliary function $f$ of eq.(\ref{heat-bath})
Moriarty \cite{Moriarty} proposed an exponential form ({\sl Mo}), 
Edwards et al.\thinspace\cite{Edward}  a Gaussian form ({\sl G}),
and Hattori and Nakajima \cite{Hattori} an ingenious hyperbolic cosine 
function ({\sl H}). 
For clarity we plotted in the fig.\thinspace\ref{O2_a}(c) these functions
together with the probability $P$ for $h=4$.

In principle the {\sl FLA}  and the {\sl AWH} algorithms follow
a similar strategy. The {\sl FLA} step function $f$ is shown in
fig.\thinspace\ref{O2_a}(a) for $9$ bins.
As explained before one should choose the optimal number $n$ for the
intervals to minimize the consumption time. This is shown for the {\sl FLA}
for three different temperatures for an anti--ferromagnet on the
three-dimensional triangular lattice in fig.\thinspace\ref{O2_a}(b).
With the choice of $n=56$ bins the Boltzmann
weight is sufficiently well approximated by an error of $15\%$ and
this choice will be almost independent on the lattice type.

Before looking at fig.\thinspace\ref{O2_a}(d) where the
Metropolis procedure is compared to the more efficient {\sl FLA} procedure,
it is interesting to compare all the different algorithms just reviewed. Simulations
are made for the two-dimensional triangular lattice (2t) and for the spin glass with $\pm J$
interactions on a four-dimensional cubic lattice (4sg). The results are presented
in fig.\thinspace\ref{O2_b} for a system size of $L=12$ for the 2t--lattice and for
$L=4$ for the 4sg--lattice.

In fig.\thinspace\ref{O2_b}(a) where the acceptance rate is displayed, one observes
that the {\sl FLA} has the best acceptance rate followed by the Hattori
algorithm. However, the consumption time displayed in fig.\thinspace\ref{O2_b}(b)
is the fundamental parameter for a comparison.
Then the situation changes dramatically. The {\sl FLA} is still
the most efficient of all algorithms with a gain of $60\%$ compared to the Gaussian 
algorithm $G$ and $200\%$ compared to the Metropolis one {\sl Me}
at the critical temperature.
The Hattori's algorithm becomes the worst of all
algorithms (at least for $T\geq T_c=0.514$, see \cite{loison_tri_2D}) 
in spite of a very good acceptance rate. This is due to a very long time of simulation
of one MC step since the function $f$ and the cumulative probability are too
complicated. This criticism is in agreement with \cite{Pawig}.

Fig.\thinspace\ref{O2_b}(c) shows the results at the critical temperature
for the Metropolis {\sl Me} and {\sl Me}$_\delta$
and for the {\sl FLA} algorithms.
These algorithms are combined with the over--relaxation
algorithm \cite{Overrelaxation} where the number of these additional updates is
denoted by {\sl OR}. The combination allows
a gain of a factor two for the {\sl FLA}
and three for the {\sl Me}. The gain in using {\sl FLA} instead of {\sl Me} is 3
without an over--relaxation step but it decreases to almost 2 with one or two such
{\sl OR}--steps which consume very little additional time.
The best number of over--relaxation
steps depends on the temperature or the value of the field $h$ and on the
lattice size. 

Fig.\thinspace\ref{O2_b}(d) displays the consumption time for a spin glass
on a four-dimensional cubic lattice (4sg) with $\pm J$ interaction
\cite{Spin_glasse_O2_4d}.  In combination with the local algorithm
the exchange algorithm \cite{exchange} is used.  There is almost no
difference between the heat bath algorithms {\sl FLA}, {\sl G},
{\sl Mo} and {\sl H}, compared to the (2t) case shown in 
fig.\thinspace\ref{O2_b}(b).
Also the Metropolis procedure is ``less worse'' when used  in combination
with the exchange algorithm. It is ``only'' 1.5 times less efficient
than the {\sl FLA} at the critical temperature shown by the squares.
This factor grows until
2 times for {\sl Me}$_{\,\delta}$ or 3 times for {\sl Me}
at lower temperatures, 
one is interested in for the spin glasses.
The best simulation
technique should  use the {\sl FLA} procedure in combination with
the over--relaxation and the exchange algorithm.

As discussed before the results do not vary strongly as
function of the size. Further results for the three-dimensional
antiferromagnetic stacked triangular lattice (3t)
and for the two-dimensional
ferromagnetic square lattice (2s) can be found in our
homepage \cite{homepage}. They are practically the same,
if taken as a function of the
relevant variable that is the local field $h$
(\ref{good_form2}). 

We want to come back to fig.\thinspace\ref{O2_a}(d) where the Metropolis
algorithm {\sl Me} 
is compared to the {\sl FLA} for various lattices as function of
the temperature. One notices that the {\sl FLA} is the most efficient
algorithm for all lattices tested. This result is completely general.

\subsection{Heisenberg spins, {\boldmath $N=3$}}
For Heisenberg spins as three component vectors with unit norm
the distribution function is according to (\ref{P_On}) and (\ref{Omega_O3})
\begin{eqnarray}
\label{eq_P_O3}
P(\theta,\phi)\cdot\sin\theta\, d\theta\cdot d\phi&=&
e^{h\cdot \cos\theta\,-h}\cdot\sin\theta\, d\theta\cdot d\phi \nonumber \\
&=&e^{h\cdot x-h} \cdot dx \cdot d\phi 
\end{eqnarray}
with $\phi$ varying from $-\pi$ to $\pi$, $\theta$ from $0$ to $\pi$, and
$x=\cos\theta\,$ from $-1$ to $1$.
$P$ is of the same form as (\ref{P_O1_2}) for the continuous Ising spin $-1\leq S < 1$
multiplied by an additional $d\phi$. Therefore all algorithms used
for this last model can also be applied  to the Heisenberg or $O(3)$ model.
In addition we test two other algorithms.
The first is the restricted Metropolis algorithm {\sl Me}$_{\,\delta}$ where
$x^{actual}-\delta \leq x^{new}\leq x^{actual}+ \delta$ and
the second one the direct heat bath {\sl DHB}$_2$ where instead of formula
(\ref{x_O1}) for an update (\ref{x_O1_best}) is used. 

For testing the algorithms an anti--ferromagnet on a the three-dimensional
stacked triangular lattice (3t) is selected.
In fig.\thinspace\ref{O3}(a) the probability $P(h=5)$
is shown together with the ``steps'' the {\sl FLA} procedure is using.
The Metropolis function $f_{Me}=1$ is also displayed.
In fig.\thinspace\ref{O3}(b)
the consumption time for the {\sl FLA} for different temperatures is
plotted as function of the ``error''. From the figure
the best choice is where the Boltzmann weight is approximated
by an error of $15\%$ which corresponds to $n=55$ bins.

It is interesting to look also at the acceptance rate shown in
fig.\thinspace\ref{O3}(c).
For the {\sl DHB} the acceptance rate is $1$, but not so for the {\sl DHB}$_2$
variant because of the rejection of value less than -1
using (\ref{x_O1_best}).

Fig.\thinspace\ref{O3}(d) shows the consumption time for the algorithms. 
As can be seen the best algorithm is the ``new'' heat bath {\sl DHB}$_2$.
It is 25\% more efficient than the usual heat bath {\sl DHB} and 
three times better than the Metropolis algorithms at the critical
temperature shown by the squares.

\subsection{Heisenberg {\boldmath $\phi^4$} model, {\boldmath $N=3$}}
In this section we study Heisenberg spins with a norm $y$ varying between
0 and $\infty$. A potential depending on $y$ is added so that
the Boltzmann probability of (\ref{eq_P_O3}) is changed to
\begin{eqnarray}
\label{P_O3_Phi4}
P(\theta,\phi,y)\cdot dy\cdot\sin\theta\, d\theta\cdot d\phi&=&
e^{h\cdot y \cdot \cos\theta\,-y^2-\lambda (y^2-1)^2}\cdot dy\cdot\sin\theta\,
d\theta\cdot d\phi \nonumber \\
&=& e^{h\cdot y\cdot x-y^2 - \lambda (y^2-1)^2} \cdot dy\cdot dx \cdot d\phi \ .
\end{eqnarray}
We plotted this probability for $h=2$ and $\lambda=1.3182$ in
fig.\thinspace\ref{O3_Phi4}(a). To study this model for its own
sake is an  interesting task. Here it serves as an example how to handle
at the same time two variables in the probability distribution.
We tested four algorithms. 

First we apply two single variable algorithms consecutively
for the variable $x$ and then for the variable $y$.
We try two restricted Metropolis procedures {\sl Me}$_{\,2\delta}$ and also two fast linear
algorithms {\sl FLA}$_2$.

In updating both variables $x$ and $y$ in a single
Monte Carlo step we use as a first method a restricted
Metropolis algorithm where $x^{actual}-\delta_x \leq x^{new}\leq x^{actual}+
\delta_x$
and
$y^{actual}-\delta_y \leq y^{new}\leq y^{actual}+ \delta_y$.
The consumption time is minimized by adjusting $\delta_x$ and $\delta_y$
to the temperature changes.
The second method is the Alias Walker Hasting algorithm {\sl AWH}. 
We did not use the {\sl FLA} algorithm since it became too difficult
to be implemented for two variables. In any case
the array would be almost as large as for the  {\sl AWH} algorithm since 
both coordinates $x$ and $y$ for each box must be stored.

In fig.\thinspace\ref{O3_Phi4}(b) we show the consumption time for simulating a
ferromagnet on a three-dimensional cubic lattice for different temperatures using
{\sl AWH} algorithm. The number of intervals $n_y$ for the $y$
variable is varied for $n_x=10$
to find the best value $n_y$. 
Similarly a graph could be shown for dependence of the time
on the number of bins $n_x$ for the $x$ variable. We found that the best choice is
$n_x=10$ and $n_y=20$. This corresponds to $n=n_x\cdot n_y=200$ bins.

In fig.\thinspace\ref{O3_Phi4}(c) and (d) the acceptance rate and the
consumption time are shown for all four algorithms. One sees that  {\sl AWH} is the
best one but almost equivalent in efficiency to the two
fast linear algorithms used
consecutively {\sl FLA}$_2$. The Metropolis algorithms {\sl Me}$_\delta$
and {\sl Me}$_{2\delta}$
are far less efficient.

In conclusion it seems not worth to consider both variables $x$ and
$y$ at the same time and it is better to update them consecutively.
However, this conclusion may hold only if the probability has a form similar to
the one in fig.\thinspace\ref{O3_Phi4}(a). The situation will be
different if, for example, there exists two or more peaks in the probability
distribution connected by regions of low probability. In this case the {\sl AWH}
algorithm should become much more efficient.

\subsection{{\boldmath $O(4)$} spins}
The  $O(4)$ spin is a four components vector with norm unity.
The probability distribution using (\ref{P_On}) and 
(\ref{Omega_O4}) is similar to the one
appearing in  $SU(2)$ gauge theory \cite{Creutz,Kajantie} and has the form
\begin{eqnarray}
\label{eq_P_O4}
P(\theta_2,\theta_1,\phi)\!\!&\cdot&\!\!\sin^2\!\theta_2\,d\theta_2\cdot
                         \sin\theta_1\,d\theta_1\cdot d\phi \nonumber \\
     	   &=& e^{h\cdot
\cos\theta_2 -h} \cdot \sin^2\!\theta_2\,d\theta_2\cdot
                         \sin \theta_1\,d\theta_1\cdot d\phi \nonumber\\
\label{eq_P_O4_2}
&=&e^{h\cdot x_2-h} \cdot \hbox{$\sqrt{1-x_2^2\,}$} \; dx_2 \cdot dx_1\cdot d\phi
\ ,
\end{eqnarray}
with $\phi$ varying from $-\pi$ to $\pi$, $\theta_i$ from $0$ to $\pi$,
$x_2=\cos \theta_2 $ and $x_1=\cos \theta_1 $ from $-1$ to $1$.
This probability is not integrable in a simple way
and therefore no direct heat bath {\sl DHB} can be used. 

Three types of Metropolis algorithms  are compared.
The first is the standard one for $N \geq 4$.
The four components are taken randomly using Gaussian random numbers and
then are normalized to get a unit vector. We denote this method by {\sl Me}$_G$
in the figure.\\
The second one, denoted by  {\sl Me},  we introduce here, will be
faster than the {\sl Me}$_G$ method for not  so large $N$ and equivalent
in performance to $Me_G$ for $N\rightarrow \infty$.
We choose randomly $\phi$, $x_1$ and $x_2$ and accept the last value with a
probability $\sqrt{1-x_2^2\,} \leq 1\,$.\\
The last one is the restricted Metropolis $Me_\delta$. We follow the second
{\sl Me} procedure but restrict the choice of the new value  by
$x_2^{actual}-\delta \leq x_2^{new}\leq x_2^{actual}+ \delta$.

We found three other algorithms in the literature, mainly
used in the context of the $SU(2)$ gauge theory.
The first one is the Creutz algorithm {\sl Cr}  \cite{Creutz}. 
It fits $P$ of (\ref{eq_P_O4_2}) with an exponential like the Moriarty algorithm
in the $XY$ case (see fig.\thinspace\ref{O4}(a)).\\
The second one is due to Fabricius et al.\ \cite{Fabricius},
which we call {\sl Fa}. With a  change of variables,  they
fit the probability $P$ depending on $x_2$ of ({\ref{eq_P_O4_2}) by the function 
$\sqrt{1-x_2}\cdot {\rm e}^{h\cdot x_2-h}\,$, but the price to pay 
is a more complicated algorithm.\\
The third {\sl Ke} is similar to the last one, but is slightly modified to gain in speed.
It is due to Kennedy and Pendleton \cite{Kennedy}.

We have implemented also the  {\sl FLA} procedure.
We calculate $x_2$ and determine $\phi$ and $x_1$ randomly.
In the final stage we use these values to create a new spin around
the local field $h$.
All the implementations of the methods introduced in this section are 
accessible at our homepage \cite{homepage}.

The results of the simulation are displayed in fig.\thinspace\ref{O4} for the
three-dimensional anti--ferromagnet on a stacked triangular lattice (3t).
Fig.\thinspace\ref{O4}(a) we show the probability 
and our function $f_{FLA}(x_2)$ with $n=20$ bins 
and $h=2$. $P(x_2)$ reaches its maximum at
$x_2^{max}= (\sqrt{1+4\,h^2} - 1)\big/2 h$
which is useful for the application of formula (\ref{equ_f_i_max}).
In the same figure the function of Creutz and the Metropolis probability
are also shown. \\
The fig.\thinspace\ref{O4}(b) shows that an error of $15\%$ corresponding to $n=55$
bins is the best choice for the {\sl FLA}.
The acceptance rate is shown in fig.\thinspace\ref{O4}(c)
and in fig.\thinspace\ref{O4}(d) the consumption time
of the various algorithms. Again the {\sl FLA} algorithm is the most
efficient algorithm. At the critical temperature shown by the squares
the gain is a factor $5$
compared to the $Me$ or $Me_\delta$ and reaches even $9$ in comparison
to the standard $Me_G$.

For the $SU(2)$ gauge theory we do not have to update $\phi$ and $x_1$,
therefore the situation is even more in favor for a simulation with the {\sl FLA}. 
In the interesting region $h\approx 1/16\,$ \cite{Kennedy},
the gain is of 30\% compared to
Kennedy's algorithm {\sl Ke}, $80\%$ to Fabricius's algorithm {\sl Fa},
and 120\% to the algorithm of Creutz {\sl Cr}.

Since the algorithms are local, the results are valid for different lattices
and sizes.

\subsection{{\boldmath $O(N)$} spins}
The $O(N)$ spins, that is vectors of unit norm in $N$ dimensions, 
do not have an experimental realization, but
can be helpful for a better understanding of the nature of phase
transitions \cite{Loison_Triangular_O6,Loison_Diep}.
The probability can be written as
\begin{eqnarray}
\label{eq_P_ON}
P(\theta_N,\cdots,\theta_1,\phi)\!\!&\cdot&\!\!\sin^{N-2}\! \theta_{N-2}\,d\theta_{N-2}\cdots
                         \sin\theta_1\,d\theta_1\cdot d\phi \nonumber \\
   	   &=& e^{h\cdot \cos\theta_{N-2}\, - h}
		   \cdot\sin^{N-2}\! \theta_{N-2}\, d\theta_{N-2}\cdots
                         \sin\theta_1\,d\theta_1\cdot d\phi \nonumber \\
\label{eq_P_ON_2}
&=&e^{h\cdot x_{N-2}\,-h} \cdot (1-x_{N-2}^2)^{\frac{N-3}{2}}\, dx_{N-2}
\cdots dx_1 \cdot  d\phi
\end{eqnarray}
with $\phi$ varying from $-\pi$ to $\pi$, $\theta_i$ from $0$ to $\pi$,
$x_i=\cos(\theta_i)$ from $-1$ to $1$.

The probability for $O(N)$ spins is not amenable to the direct heat
bath {\sl DHB} technique. This problem is similar to the $O(4)$ case 
with the probability given by (\ref{eq_P_O4_2}). In addition for the $O(N)$
case the variables $\{x_{N-3},\cdots,x_2\}$ can also not be dealt with 
in a simple way. See fig.\thinspace\ref{O6_a}(a--c) where
for $N=6$ the probabilities are shown for $x_4$, $x_3$ and $x_2$.

We have used the same Metropolis algorithms as in the previous
section, that is the simple Metropolis {\sl Me}, the restricted Metropolis
{\sl Me}$_\delta$, and the standard one $Me_G$. For the two first algorithms
{\sl Me} and {\sl Me}$_\delta$, the $\{x_{N-3},\cdots,x_2\}$ variables are
determined using a rejection method: a random number $x_n$ between -1 
and 1 is accepted as $x_n^{new}$ with a probability $(1-x_n^2)^{\frac{n-1}{2}}$.
The refusal rate is proportional to the area between the curve
``sinus'' in fig.\thinspace\ref{O6_a}(b,c) and the curve
$(1-x_n^2)^{\frac{n-1}{2}}$. The $x_{N-2}$ is calculated randomly between
-1 an 1 (or between $x^{actual}\pm \delta$) and accepted with
Hasting formula (\ref{Hasting}) with $f=1$.

In analogy to Moriarty's algorithm for $XY$ spin, to the  heat
bath for Heisenberg spin, or to the algorithm of  Creutz for $O(4)$ spins, we
introduce an exponential algorithm {\sl Ex} for the $x_{N-2}$
variable. We calculate $x_{N-2}$
using $x_{N-2}= - 1 + \frac{1}{h}\,\log[1+ \ran\cdot (e^{2 h} -1) ] $
and accept it with a probability $(1-x_{N-2}^2)^{\frac{N-3}{2}}$.
The others $\{x_n\}$ are calculated in a similar way as for {\sl Me}.

We want to show that it is profitable to use the {\sl FLA}. For $N=6$ the best choice
corresponds to 15\% of error and $n=55$ bins for $x_{4}$ as shown in 
fig.\thinspace\ref{O6_a}(d). For $x_3$ and $x_2$ the
{\sl FLA} is also used with 200 bins (see fig.\thinspace\ref{O6_a}(b) and (c) with 10
and 20 bins, respectively).  The probability $P(x_{N-2})$ reaches its maximum at
$x_{N-2}^{max}= \bigl(\sqrt{(N-3)^2+4\,h^2}\, - (N-3)\bigr)\big/2h$
which is helpful for constructing the step function $f$ for the
{\sl FLA} with eq.(\ref{equ_f_i_max}). 

In fig.\thinspace\ref{O6_b}(a) and (b) we put the essential results for $N=6$,
obtained for an anti--ferromagnet on a three-dimensional stacked 
triangular lattice (3t)\cite{Loison_Triangular_O6,Loison_Diep}.
In fig.\thinspace\ref{O6_b}(a) the acceptance rate for the different
algorithms is shown. We want to point to the differences
between Hasting's method (e.g.\ Metropolis) and the heat bath
methods (e.g.\ {\sl FLA} and {\sl Ex}). In the first case, if the new spin is
not accepted, all values $x_n$ and the angle $\phi$ are discarded.
In the other case for the heat bath method one always gets a new state.
Therefore the last algorithm becomes much more efficient if the number
of values $x_n$ to be calculated increases.
In fig.\thinspace\ref{O6_b}(b) one can compare the consumption time for the different algorithms.
One sees, that {\sl FLA} is the most efficient of all the algorithms tested.

Fig.\thinspace\ref{O6_b}(c) and (d) contains a summary of results for 
different spin dimension $1\leq N \leq 6$, that is the acceptance rate and
the consumption time at the critical temperature for an anti--ferromagnet
on the 3t--lattice. The algorithms compared are the fast linear algorithm {\sl FLA},
the Metropolis algorithm {\sl Me}, the restricted Metropolis
algorithm {\sl Me}$_\delta$ and the Gaussian Metropolis algorithm {\sl Me}$_G$.
One observes that the {\sl FLA} is more efficient than any Metropolis 
algorithms and that the advantage is larger for an increasing number of spin
components $N$. This conclusion is totally general and does not depend on
the system size nor on the lattice type.

\subsection{Other types of spin Hamiltonians \label{other_type_H}}
Until now we only have dealt  with Hamiltonians which can be
written in the form given by (\ref{good_form1}) and (\ref{good_form2}).
In fact the Fast Linear Algorithm {\sl FLA} can only be used for this kind of
Hamiltonian. However, the Alias Walker Hasting procedure is more flexible.

For example, for a Hamiltonian $H=-\sum_{<ij>}(S_i\cdot S_j)^3$ \cite{Loison_cubic}
the ground state is a conventional ferromagnetic one and at finite temperatures the spins
will still be oriented around the direction of all their neighboring spins. However, one
cannot define a local field $h$ as in (\ref{good_form2}) for this cubic exchange.
This means that the {\sl FLA} algorithm or any other heat bath method cannot work.
For these methods one needs a function
$f \geq P$ and since there is no formula of
$P(h)$ definable, one has only the inefficient choice of a constant $f$ 
equal to the maximum of $P$. 
However, Hasting's method is not limited by the condition $f \geq P$ .
Among these methods the {\sl AWH} algorithm with a choice for
$f$, for example similar to the case with linear interactions we discussed,
should offer a possibility for an efficient simulation. 


\section{Conclusion}

In this article we have tried to review as completely as possible the
available  canonical local algorithms for discrete and continuous spin systems.
Starting from the difference between Hasting's and heat bath methods
we analyzed the algorithm regarding their numerical efficiency.
In all cases except the standard Ising case,
we could demonstrate that  by modifying known algorithms
or to construct new ones the efficiency increases compared
to the standard ones, in particular compared to the widely known and
used Metropolis algorithm. 

For discrete spins (Ising model, Blume-Capel model, Potts model and clock
model) the best algorithm is a restricted heat bath 
Walker Hasting algorithm where
the new state is chosen by the heat bath procedure among all accessible states
excluding the actual one. Then Hasting's method is used to accept or
reject the new state.
To accelerate the simulation we have proposed
Walker's alias method, usually not used in this context.

For continuous spins (continuous Ising spin, $XY$ spins, Heisenberg
spin and $O(N\geq4)$ spins) we introduced different Walker Hasting's methods
and the Fast Linear Algorithm. These methods are efficient at all
temperatures and achieve a gain at the critical temperature of usually more
than two compared to the Metropolis algorithm. For low
temperature the gain becomes even larger. 

We use arrays heavily to store data calculated prior to the simulation.
Therefore the efficiency of the introduced algorithms depend
strongly on the amount of fast cache memory of the processor.
Since this amount will certainly increase in the future, 
the algorithms proposed should become even more efficient than quoted in
this article.

We tested the algorithms on a number of lattices varying the sizes
but the conclusion we reach are valid whatever the lattices
and sizes are since local algorithms are compared.
Moreover we think that these improvements of the algorithms applied to spin systems
are of general nature and could be applied also outside of this field.

\section {Acknowledgments}
We are grateful to Flavio Nogueira 
and Sonoe Sato for their support,
discussions, and for  reading the manuscript critically.


\vskip 1.5cm
\leftline{\bf Figures \& Figure Captions}
\newpage
\pagestyle{empty}
\setlength{\textwidth}{14cm}
\renewcommand{\baselinestretch}{1}
%
\vskip 0.5cm
\begin{figure}[t]
\centerline{
\psfig{figure=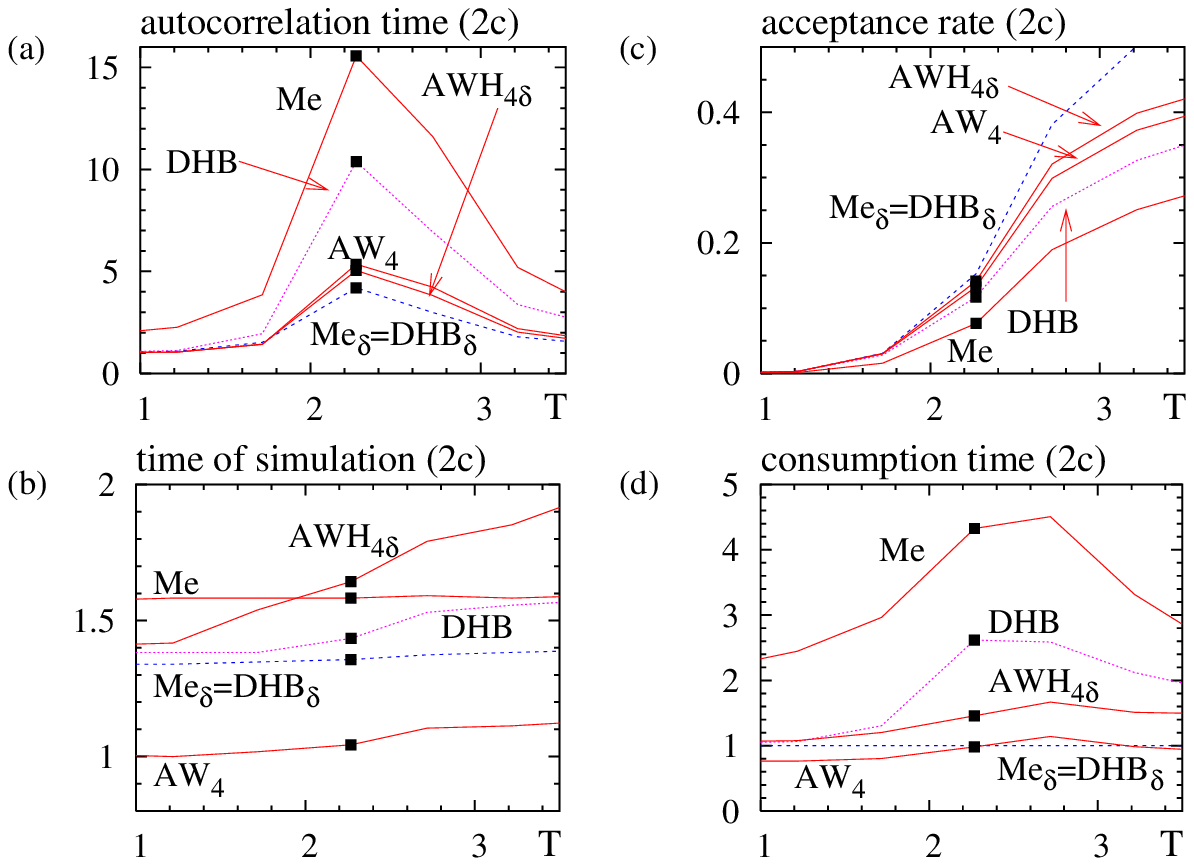,width=9.0cm}
}
\vskip -0.2cm
\caption{
\label{Ising}
Ising model. Results for a ferromagnet on a square lattice.
}
\end{figure}
%
\vskip -0.3cm
\begin{figure}
\centerline{
\psfig{figure=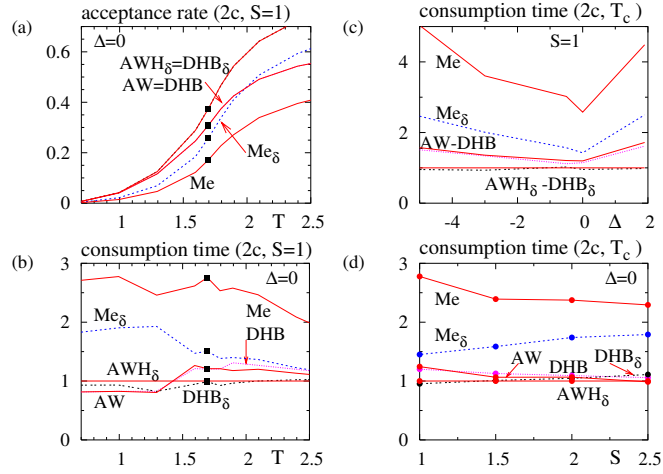,width=9.0cm}
}
\vskip -0.2cm
\caption{
\label{Ising_S_a}
Blume-Capel model  on a square lattice, see text for explanations.
}
\end{figure}
%
\vskip 0.3cm
\begin{figure}
\centerline{
\psfig{figure=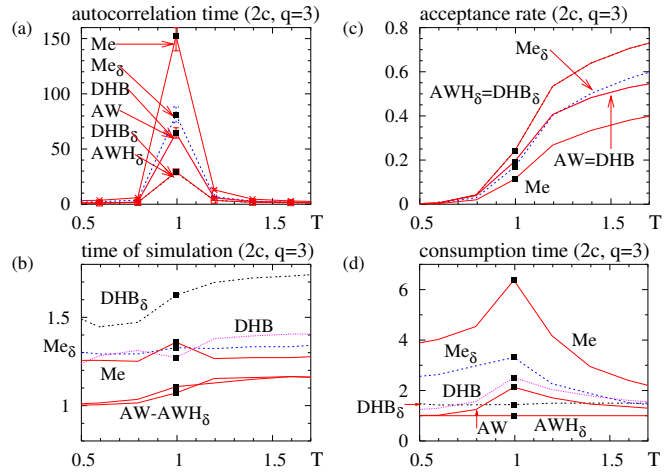,width=9.0cm}
}
\vskip -0.2cm
\caption{
\label{Potts_a}
Three state Potts model. Results for a ferromagnet on a square lattice.
}
\end{figure}
%
\vskip 0.3cm
\begin{figure}
\begin{minipage}{13cm}
\centerline{
\psfig{figure=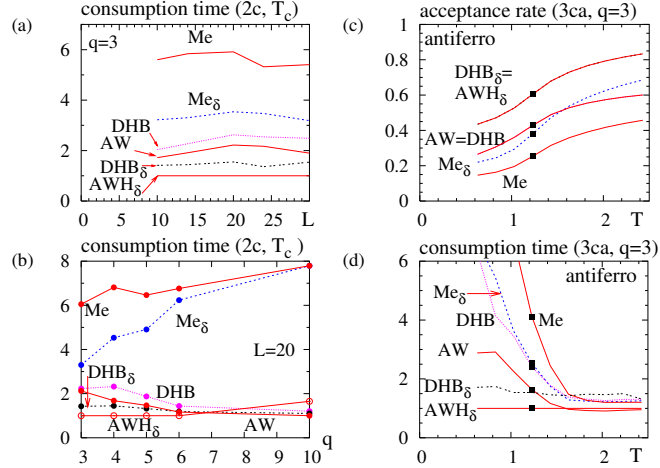,width=9.0cm}
}
\vskip -0.2cm
\caption{
\label{Potts_b}
$q$ state Potts model. Results for a ferromagnet on square lattice (2s)
and for an anti--ferromagnet on cubic lattice (3ca). $L$ is the system size.
}
\end{minipage}
\end{figure}
%
\begin{figure}
\begin{minipage}{13cm}
\centerline{
\psfig{figure=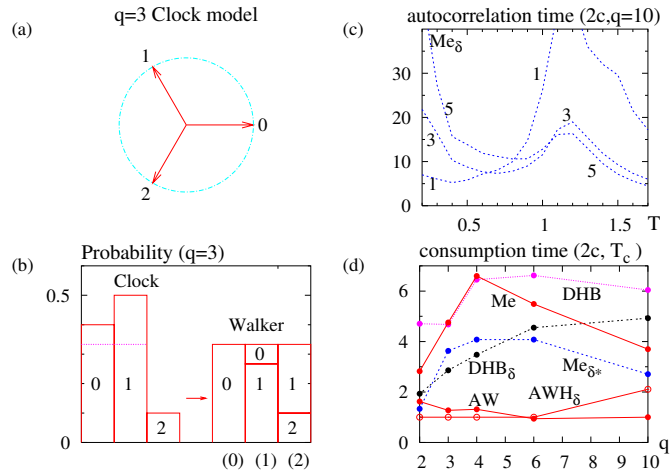,width=9.0cm}
}
\vskip -0.2cm
\caption{
\label{Clock_a}
$q$ state clock model: (a) model, (b) probability and Walker's ``boxes'',
(c) and (d) show results for a ferromagnet on a square lattice (2s).
}
\end{minipage}
\end{figure}
%
\vskip 0.5cm
\begin{figure}[b]
\centerline{
\psfig{figure=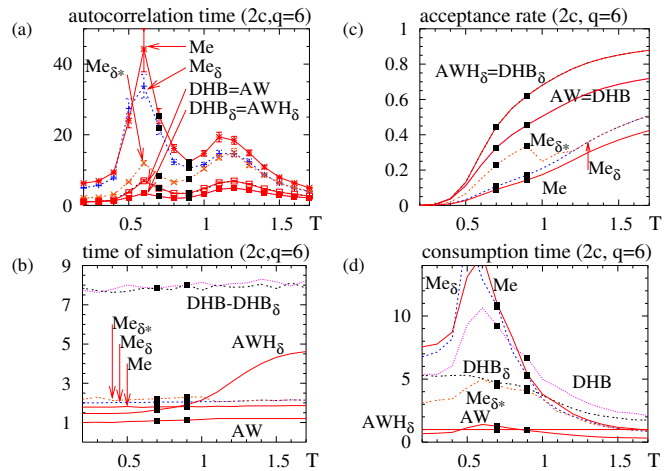,width=9.0cm}
}
\vskip -0.2cm
\caption{
\label{Clock_b}
$q = 6$ state clock model. Results for a ferromagnet on a square lattice.
}
\end{figure}
%
\vskip -0.3cm
\begin{figure}[t]
\begin{minipage}{13cm}
\centerline{
\psfig{figure=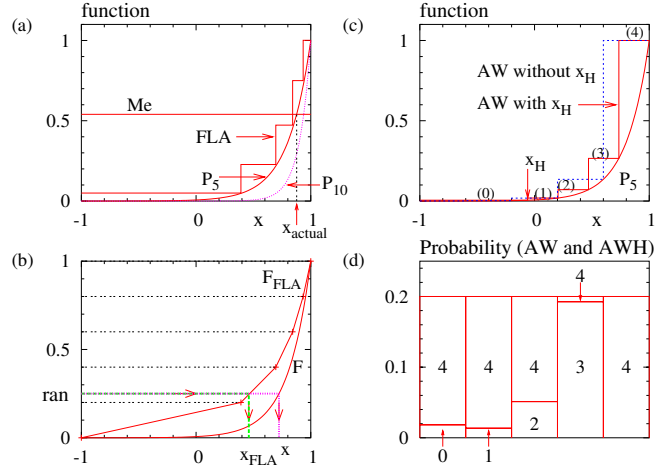,width=9.0cm}
}
\vskip -0.2cm
\caption{
\label{O1_a}
Continuous Ising spin $-1\leq S < 1$:
(a) \& (c) probability $P$ and step functions $f$,
(b) cumulative probability of $P$ ($F$) and $f_{FLA}$ ($F_{FLA}$)
with equidistant dashed lines,
(d) Walker's boxes corresponding to ``{\sl  AW} with $x_H$'' in (c).
}
\end{minipage}
\end{figure}
%
\vskip -0.5cm
\begin{figure}
\begin{minipage}{13cm}
\centerline{
\psfig{figure=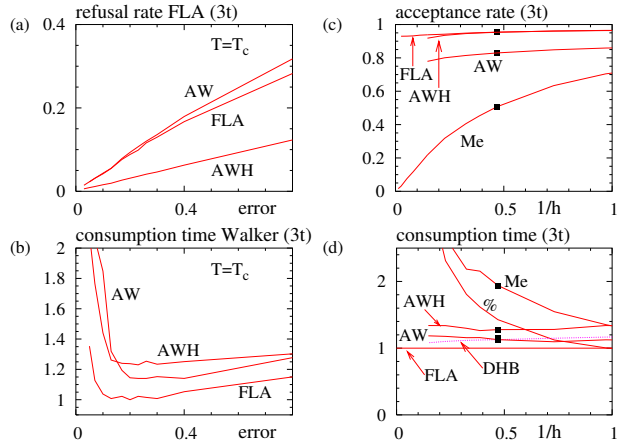,width=8.5cm}
}
\vskip -0.2cm
\caption{
\label{O1_b}
Continuous Ising spin $-1\leq S < 1$. Results for an anti--ferromagnet
on a 3d  stacked triangular lattice (3t), see text for explanations.
}
\end{minipage}
\end{figure}
\vskip 0.5cm
\begin{figure}
\begin{minipage}{13cm}
\centerline{
\psfig{figure=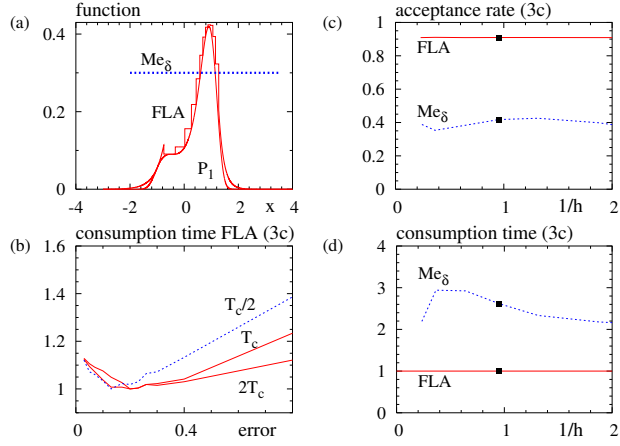,width=8.5cm}
}
\vskip -0.5cm
\caption{
\label{O1_Phi4}
Ising $\phi^4$ model. (a) probability $P$ and function $f_{FLA}$,
(b--d) results for a ferromagnet on a cubic
lattice (3c) discussed in the text.
}
\end{minipage}
\end{figure}
\vskip -0.5cm
\begin{figure}
\begin{minipage}{13cm}
\centerline{
\psfig{figure=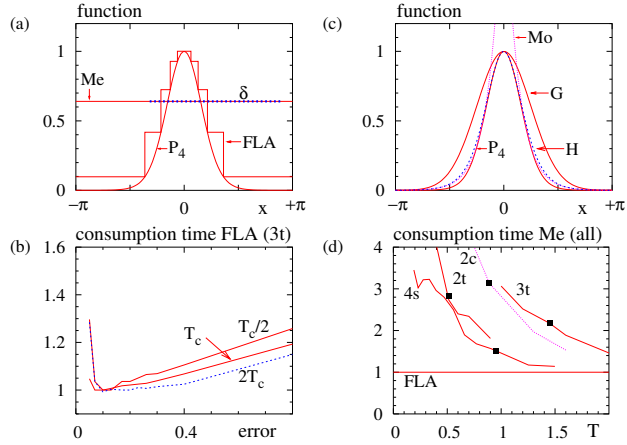,width=8.5cm}
}
\vskip -0.2cm
\caption{
\label{O2_a}
$XY$ spins. (a) \& (c) probability $P$ and functions $f$,
consumption time (b) for {\sl FLA}  and (d) for Metropolis  compared to {\sl FLA}
for different lattices.
}
\end{minipage}
\end{figure}
\vskip 0.5cm
\begin{figure}
\begin{minipage}{13cm}
\centerline{
\psfig{figure=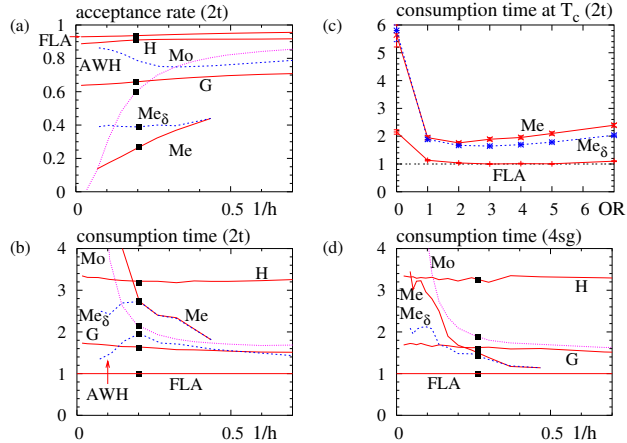,width=8.5cm}
}
\vskip -0.2cm
\caption{
\label{O2_b}
$XY$ spins. (a) \& (b) results for an anti--ferromagnet on a 2d triangular
lattice (2t), (c) consumption time together with over--relaxation steps {\sl OR},
and (d) for a 4d spin glass (4sg) in combination with the exchange algorithm.
}
\end{minipage}
\end{figure}
\vskip 0.5cm
\begin{figure}
\centerline{
\psfig{figure=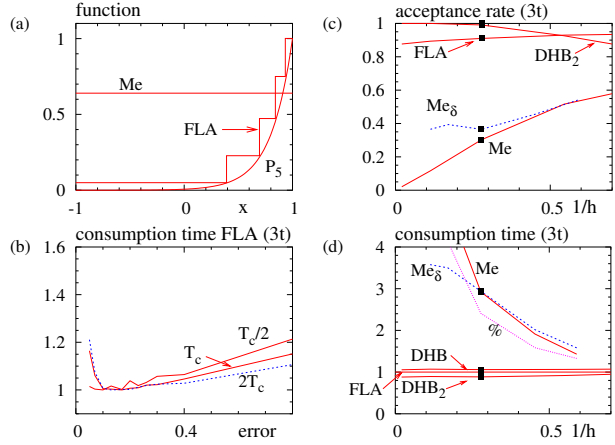,width=8.5cm}
}
\vskip -0.2cm
\caption{
\label{O3}
Results for a Heisenberg  anti--ferromagnet on a 3d triangular lattice.
}
\end{figure}
\vskip 0.5cm
\begin{figure}
\begin{minipage}{13cm}
\centerline{
\psfig{figure=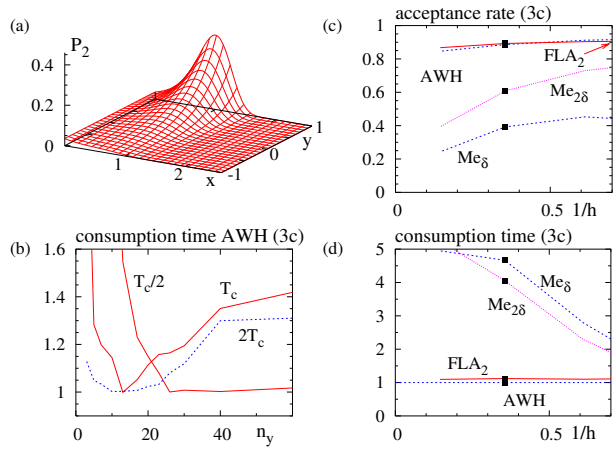,width=8.5cm}
}
\vskip -0.2 cm
\caption{
\label{O3_Phi4}
Heisenberg $\phi^4$ model:
(a) the probability $P$ depending on $x= \cos \theta$ and the norm $y$,
(b--d) results for a ferromagnet on a cubic lattice.
}
\end{minipage}
\end{figure}
\vskip 0.5cm
\begin{figure}
\centerline{
\psfig{figure=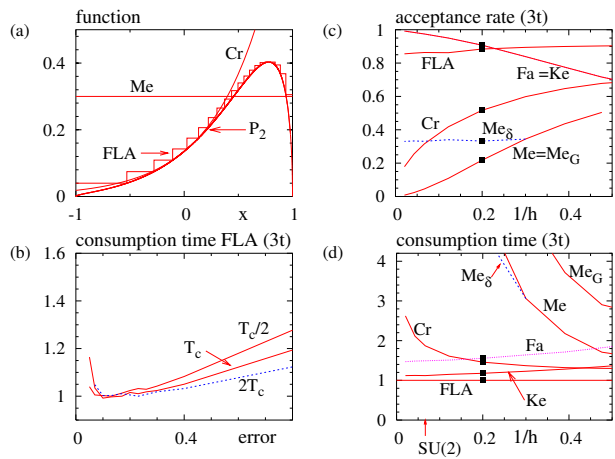,width=8.5cm}
}
\vskip -0.2cm
\caption{
\label{O4}
Four components spins $O(4)$.
Results for the 3d anti--ferromagnetic on a triangular
lattice.
}
\end{figure}
\vskip 0.5cm
\begin{figure}
\begin{minipage}{13cm}
\centerline{
\psfig{figure=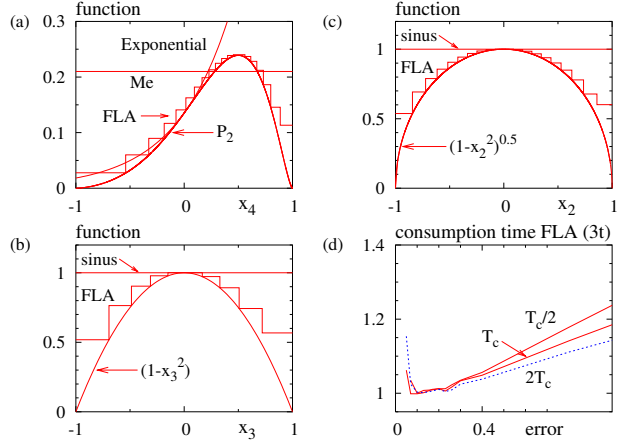,width=8.5cm}
}
\vskip -0.2cm
\caption{
\label{O6_a}
Six component spins $O(6)$:
(a) (b) and (c) probabilities and the corresponding step functions $f$.
(d) Search for a minimal consumption time.
}
\end{minipage}
\end{figure}
\vskip 0.5cm
\begin{figure}
\begin{minipage}{13cm}
\centerline{
\psfig{figure=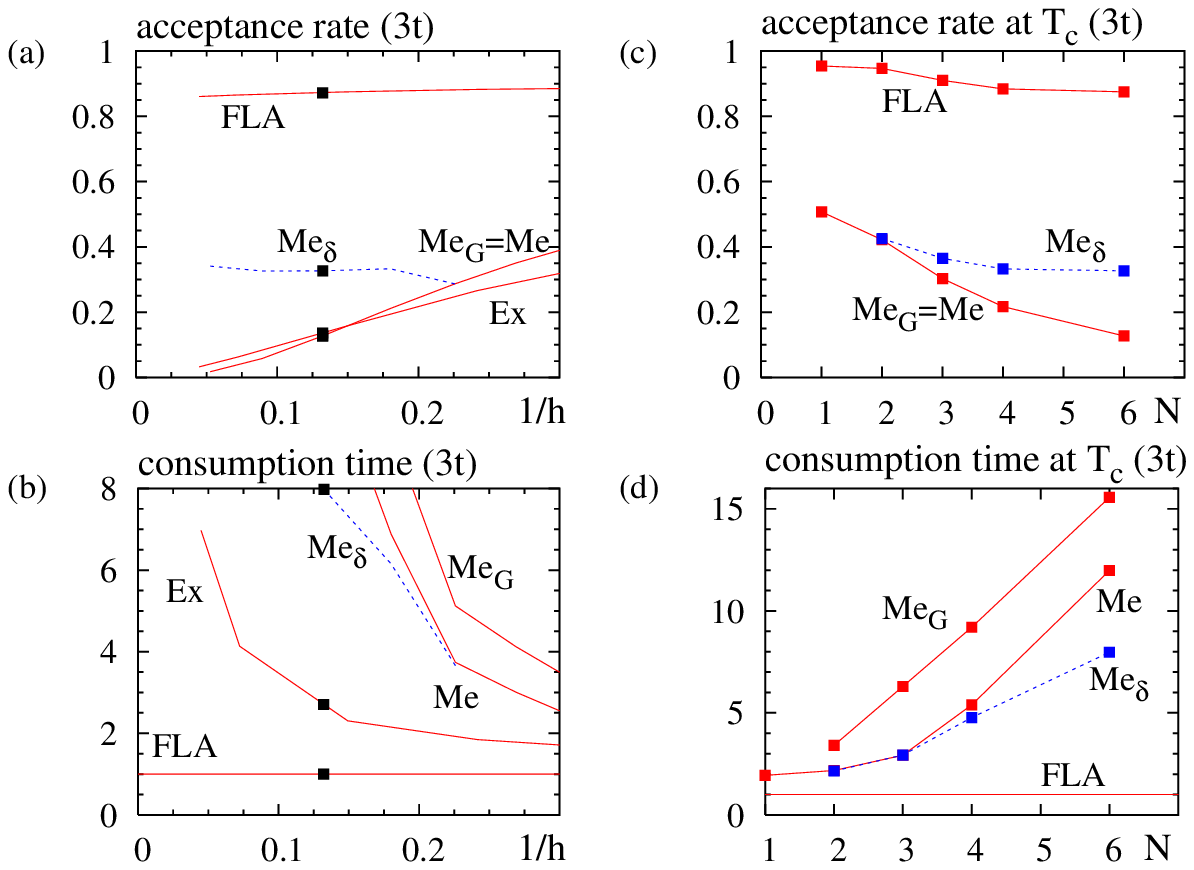,width=8.5cm}
}
\vskip -0.2 cm
\caption{
\label{O6_b}
$N$ components spins.
Results for an  anti--ferromagnet on a 3d triangular
lattice,
(a) \& (b) for $N=6$.
(c) \& (d) for different $N$ at $T_c$.
}
\end{minipage}
\end{figure}

\begin{thebibliography}{99}
%
\bibitem {Metropolis} N. Metropolis, A.W. Rosenbluth, M.N. Rosenbluth,
A.H. Teller and E. Teller, J. Chem. Phys. {\bf 21}, 1087 (1953).
%
\bibitem {homepage}{\verb#http://www.physik.fu-berlin.de/~loison/#}
%
\bibitem {Plascak} J.A. Plascak and D.P. Landau, Phys. Rev. E {\bf 67}, 015103 (2003).
%
\bibitem {exchange}  K. Hukushima and K. Nemoto, J. Phys. Soc. Jap.{\bf 65},
1604 (1996), Phys. Rev. E {\bf 61} R1008 (2000).
\bibitem{Overrelaxation} M. Creutz, Phys. Rev. D {\bf 36}, 515 (1987).
%
\bibitem {Berg} B. Berg and T. Neuhaus, Phys. Lett. B {\bf 267}, 249 (1991),
Phys. Rev. Lett. {\bf 68}, 9 (1992).
%
\bibitem {Loison_Simon} D. Loison and P. Simon,  Phys. Rev. B {\bf 61}, 6114 (2000).
%
\bibitem {Gentle} J.E. Gentle, {\sl Random number Generation and Monte
Carlo Methods}, Springer, 1998.
%
\bibitem {Hasting} W.K. Hasting, Biometrika {\bf 170}, 97 (1970).
%
\bibitem {Walker}  A. J. Walker, ACM Transaction on Mathematic
Software {\bf 3}, 253 (1977).
%
\bibitem {Peterson} R. Kronmal and A. Peterson, American Statistician,
{\bf 33}, 214 (1979).
%
\bibitem {Blume} M. Blume, Phys. Rev. {\bf 141}, 517 (1966); H.W. Capel,
Physica {\bf 32}, 966 (1966).
%
\bibitem {Blume1.5} e.\ g.:
J. A. Plascak and D.P. Landau, Phys. Rev. E {\bf 67}, 151103 (2003);\\
R. da Silva, N.A Alves, and J.R.D. de Felicio, Phys, Rev. E {\bf 67}, 57102 (2003);\\
S. Grollau, Phys. Rev. E {\bf 65}, 56103 (2002);\\
M.M. Tsypin and H.W.J. Bl{\"o}te, Phys. Rev. E {\bf 62}, 73 (2000);\\
S. Bekhechi and A. Benyoussef, Phys. Rev. B {\bf 56}, 13954 (1997).
%
\bibitem {Potts} R.B. Potts, Proc. Cambridge Philos. Soc. {\bf 48}, 106 (1952).
%
\bibitem {Potts_today} e.\ g.: Z. F. Wang, and 
B.W. Southern, Phys. Rev. B {\bf 68}, 94419 (2003); ibid {\bf 67}, 54415 (2003);\\
M. Itakura, Phys. Rev. B {\bf 60}, 6558 (1999);\\
G. Pal{\'a}gyi, C. Chatelain, B. Berche, et al., Eur. Phys. J. B {\bf 13}, 357 (2000);\\
M. Reuhl, P. Nielaba, and K. Binder, Eur. Phys. J. B {\bf 2}, 225 (1998).
%
\bibitem {LandauClock} M.S.S. Challa and D.P. Landau, Phys. Rev. B {\bf 33} 437 (1986);\\
W. Scott and H.L. Scott, J. Phys. A: Math. Gen. {\bf 22} 4463 (1989).
%
\bibitem {Clock}  e.\ g.:
J. Hove and A. Sudbo, Phys. Rev. E {\bf 68}, 46107 (2003);\\
A. Benyoussef, M. Loulidi, and A. Rachadi, Phys. Rev. B {\bf 67}, 094415 (2003);\\
N. Todoroki, Y. Ueno, and S. Miyashita, Phys. Rev. B {\bf 66}, 214405 (2002);\\
J. D. Noh, H. Rieger, M. Enderle, and K. Knorr, Phys. Rev. E {\bf 66}, 026111 (2002);\\
F. J. Resende and B. V. Costa, Phys. Rev. E {\bf 58}, 5183 (1998).
%
\bibitem {Miyatake} Y. Miyatake, M. Yamamoto, J.J. Kim, 
M. Toyonaga and O. Nagai, J. Phys. C: Solid State Phys. {\bf 19}, 2539 (1986).
%
\bibitem {Bayong} E. Bayong and H. T. Diep, Phys. Rev. B {\bf 59}, 11919 (1999).
%
\bibitem {Hasenbusch}   M. Hasenbusch, K. Pinn, S. Vinti, Phys. Rev. B
{\bf 59} 11471  (1999) and references therein.
%
\bibitem {Creutz_Jacob} M. Creutz, L. Jacob, and C. Rebbi, Phys. Rep. 95, 201 (1983).
%
\bibitem {Moriarty} K.J. Moriarty, Phys. Rev. D {\bf 25}, 2185 (1982).
%
\bibitem {Edward} R.G. Edwards, J. Goodman, A.D. Sokal, Nucl. Phys. B {\bf 354},
289 (1991).
%
\bibitem {Hattori} T. Hattori and H. Nakajima, Nucl. Phys. B (Proc. Suppl.) {\bf 26},
635 (1992), J. of Comp. Phys. {\bf 121}, 238 (1995).
%
\bibitem {loison_tri_2D} D. Loison, \verb#http://www.physik.fu-berlin.de/~loison/articles/reference20.html#
%
\bibitem {Pawig} S.G. Pawig, K. Pinn, Int. J. Mod. Phys. C {\bf 9}, 727 (1998).
%
\bibitem {Spin_glasse_O2_4d}  H.G. Katzgraber and A.P. Young, Phys. Rev. B {\bf 65},
214401 (2002).
%
\bibitem {Creutz} M. Creutz, Phys. Rev. D {\bf 21}, 2308 (1980).
%
\bibitem {Kajantie} K. Kajantie, M. Laine, K. Rummukainen, M. Shaposhnikov,
Nucl. Phys. B {\bf 466}, 189 (1996).
%
\bibitem {Fabricius} K. Fabricius and O. Haan, Phys. Lett. B {\bf 143}, 459 (1984).
%
\bibitem {Kennedy} A.D. Kennedy and B.J. Pendleton, Phys. Lett. B {\bf 156}, 393 (1985).
%
\bibitem {Loison_Triangular_O6}  D. Loison, A.I. Sokolov, B. Delamotte,
S.A. Antonenko, K.D. Schotte and H.T. Diep, JETP Letters {\bf 76}, 337 (2000).
%
\bibitem {Loison_Diep}  D. Loison in {\sl Frustrated spin systems},
ed.\ H.T. Diep, World Scientific,  2nd edit.\ (2004).
%
\bibitem {Loison_cubic}  D. Loison, Phys. Lett. A {\bf 257}, 83 (1999);
ibid. {\bf 264}, 208 (1999).
%
\end{thebibliography}
\end{document}